\documentclass[a4paper,11pt]{article}
\usepackage{jheppub} % for details on the use of the package, please
\usepackage[T1]{fontenc} % if needed
\usepackage{amsmath}
\usepackage{graphicx}
\usepackage{array,multirow}
\usepackage{url}
\usepackage{fancyhdr}
\usepackage{lastpage}
\usepackage{tabularx}
\usepackage{rotating}
\usepackage{slashed}
\def\be{\begin{eqnarray}}
\def\ee{\end{eqnarray}}
\newcommand{\beq}{\begin{eqnarray}}
\newcommand{\eeq}{\end{eqnarray}}
\newcommand{\mt}[1]{\textrm{\tiny #1}}

\def\R{{\mathcal{R}}}

\def\O{{\mathcal{O}}}

\def\z{{z}}
\def\Leff{{L_{\rm eff}}}

\def\phiH{{\phi_H}}
\def\phiM{\phi_M}

\def\LIR{L_{\rm IR}}
\def\mIR{{m^2_{\rm IR}}}

\def\GT{{G^{\rm {(T)}}}}

\def\Zm{\mathcal{Z}_{\rm aniso}}

\def\Zphi{\mathcal{Z}_{\rm bulk}}

\def\Zt{{\mathcal{Z}_0}}

\def\phit{\tilde \phi}

\def\Llphi{{\mathcal{L}_{\rm bulk}}}
\def\Lrphi{{\mathcal{R}_{\rm bulk}}}

\def\Llm{{\mathcal{L}_{\rm aniso}}}
\def\Lrm{{\mathcal{R}_{\rm aniso}}}

\newcommand{\eq}[1] {eq.~(\ref{#1})}
\newcommand{\fig}[1] {Fig.~\ref{#1}}

\newcommand{\eqq}[1]{(\ref{#1})}

\title{Thermodynamics, transport and relaxation in non-conformal theories}
\arxivnumber{1603.01254}

\author[a,c]{Maximilian Attems,}
\author[a,b]{Jorge Casalderrey-Solana,}
\author[c,d]{David Mateos,}
\author[e]{Ioannis Papadimitriou,}
\author[f]{Daniel Santos-Oliv\'an,}
\author[f]{Carlos F. Sopuerta,}
\author[c]{Miquel Triana,}
\author[c]{Miguel Zilh\~ao,}

\affiliation[a]{
Departament d'Estructura i Constituents
de la Mat\`eria and Institut de Ci\`encies del Cosmos (ICCUB),
Universitat de Barcelona, Mart\'\i \ i Franqu\`es 1, 08028 Barcelona, Spain}
\affiliation[b]{
Rudolf Peierls Centre for Theoretical Physics, University of Oxford, 1 Keble Road, Oxford OX1 3NP, United Kingdom}
\affiliation[c]{
 Departament de F\'isica Fonamental and Institut de Ci\`encies del Cosmos (ICCUB),
Universitat de Barcelona, Mart\'\i \ i Franqu\`es 1, 08028 Barcelona, Spain}
\affiliation[d]{
Instituci\'o Catalana de Recerca i Estudis Avan\c{c}ats (ICREA),
Passeig Llu\'is Companys 23, ES-08010, Barcelona, Spain.}
\affiliation[e]{SISSA and INFN - Sezione di Trieste, Via Bonomea 265, I 34136 Trieste, Italy}
\affiliation[f]{
Institut de Ci\`encies de l'Espai (CSIC-IEEC), Campus UAB,
Carrer de Can Magrans s/n, 08193 Cerdanyola del Vall\`es, Spain}

\emailAdd{attems@icc.ub.edu}

\emailAdd{jorge.casalderreysolana@physics.ox.ac.uk}

\emailAdd{dmateos@icrea.cat}

\emailAdd{ioannis.papadimitriou@sissa.it}

\emailAdd{santos@ieec.uab.es}

\emailAdd{sopuerta@ieec.uab.es}

\emailAdd{mtriana@ffn.ub.edu}

\emailAdd{mzilhao@ffn.ub.es}

\preprint{{\footnotesize ICCUB-16-011}, \footnotesize{SISSA 17/2016/FISI} }

\abstract{
We study the equilibrium and near-equilibrium properties of a holographic 
five-dimensional model consisting of Einstein gravity coupled to a scalar field with a non-trivial potential. The dual four-dimensional gauge theory is not conformal and, at zero temperature, exhibits a renormalisation group flow between two different fixed points. We quantify the deviations from conformality both in terms of thermodynamic observables and in terms of the bulk viscosity of the theory. The ratio of bulk over shear viscosity violates Buchel's bound. We study relaxation of small-amplitude, homogeneous perturbations by computing the quasi-normal modes of the system at zero spatial momentum. In this approximation we identify two different relaxation channels. At high temperatures, the different pressures first become approximately equal to one another, and subsequently this average pressure evolves towards the equilibrium value dictated by the equation of state.  At low temperatures, the average pressure first evolves towards  the equilibrium pressure, and only later the different pressures become approximately equal to one another. 
}

\begin{document} 
\maketitle
\flushbottom

%%%%%%%%%%%%%%%%
\section{Introduction}
\label{sec:intro}

The understanding of the out-of-equilibrium dynamics of matter is an important challenge ubiquitous at all energy scales. A particularly interesting case is the understanding of these dynamics in strongly coupled systems. Examples include strongly correlated electrons, cold atoms and the small drops of Quark-Gluon Plasma (QGP) formed in relativistic colliders such as RHIC or the LHC. The latter case motivates the study of the relaxation process in strongly coupled non-abelian field theories. The gauge/string duality provides a fascinating tool to address this problem in a wide range of theories.

The duality has already provided insights into the dynamics of strongly coupled,  deconfined, non-abelian matter of relevance for the heavy ions programme (see e.g.~\cite{CasalderreySolana:2011us} and references therein). The study of the off-equilibrium  dynamics of Conformal Field Theories (CFT), most notably $\mathcal{N}=4$ super Yang-Mills (SYM) theory, has shown  that hydrodynamics is a much better approximation to the evolution of this type of matter than ever thought before. Indeed, examples based on flow motions imposed by symmetries \cite{Heller:2011ju,Chesler:2009cy} or by explicit simulations of the collision dynamics \cite{Chesler:2015wra,Chesler:2013lia,Casalderrey-Solana:2013sxa,Casalderrey-Solana:2013aba} have shown that hydrodynamics provides a good approximation to the complete evolution of the system at time and distance scales as small as a fraction of the (local) inverse temperature of the system. This occurs even in situations in which gradient corrections to the hydrodynamic stress tensor are large, extending the applicability beyond a simple gradient expansion (see also \cite{Heller:2014wfa}). This observation has led to the coining of the term ``hydrodynamisation'' to refer to the process by which a system comes to be well described by hydrodynamics, in order to differentiate this process from (local) thermalisation.
This observation, first made in holographic computations, has now been noted in Boltzmann equation-based analysis of out-of-equilibrium dynamics, when the strength of the coupling in the collision kernel is extrapolated to large values \cite{Kurkela:2015qoa}. The success of hydrodynamics to capture the evolution of 
out-of-equilibrium matter may be at the origin of the strong collective behaviour observed in very small systems, such as Au-Au collisions at RHIC \cite{Ackermann:2000tr,Adler:2003kt,Back:2004mh}, Pb-Pb \cite{ATLAS:2012at,Chatrchyan:2012ta,Aamodt:2010pa,Adam:2016izf}, p-Pb \cite{Aad:2014lta,Khachatryan:2015waa,Abelev:2014mda} and p-p \cite{Aad:2015gqa} collisions at the LHC. The  holographic analysis of  collisions of small systems \cite{Chesler:2016ceu,Chesler:2015bba}  supports this viewpoint.  

In a CFT the vanishing of the trace of the stress tensor implies that the equation of state, namely 
\be
\bar{p} = \frac{1}{3} e \,,
\ee
where 
\be
\bar{p} = \frac{1}{3}\left( p_x + p_y + p_z \right) 
\ee
is the average pressure, is fixed by symmetry. As a consequence, the equation of state is always obeyed both in and out of equilibrium. We emphasize that the equation of state fixes only the average pressure in terms of the energy density, but not the individual pressures. For this reason the  relaxation towards equilibrium in a CFT typically involves ``isotropization'', namely the process by which the different pressures become approximately equal to one another. 

The applicability of the gauge/string duality is not restricted to CFTs. By now infinite families of non-conformal examples are known. One of the main new features in these theories as compared to their conformal cousins is that new channels exist for the relaxation of the out-of-equilibrium matter. In particular, in non-conformal theories the equation of state is not fixed by symmetry. As a consequence, out of equilibrium the energy density and the average pressure may fluctuate independently. Therefore, the relaxation towards equilibrium in these theories involves the evolution of the energy density and the average pressure towards asymptotic values related to one another by the equation of state (EoS). When this happens we will say that the system has ``EoSized''  and we will refer to this process as ``EoSization''. 

Another important motivation for studying non-conformal theories is the connection with hot Quantum Chromodynamics (QCD) and heavy ion collisions. As is well known, QCD is a non-conformal theory even in the limit of vanishing quark masses. State-of-the-art determinations of the QCD equation of state via lattice QCD \cite{Borsanyi:2013bia,Bazavov:2014pvz} show that, in equilibrium, the trace of the stress tensor normalised by the enthalpy attains values of order one close to the QCD transition. At high temperature this ratio quickly approaches zero, indicating that QCD behaves as an almost-conformal theory in this regime. However, the experimental exploration of the QCD phase diagram via 
high-energy heavy ion collisions can only reach temperatures a few times larger than the critical temperature. Even though most central, top-energy LHC collisions lead to initial temperatures well into the quasi-conformal regime, the subsequent evolution and cooling of the QGP after production spans all temperature regimes, including those in which non-conformal effects are maximal. In fact, recent attempts for high-precision extraction of the shear viscosity of the QGP have highlighted the need to include the bulk viscosity of the plasma, which is a purely non-conformal effect \cite{Ryu:2015vwa}. 
Furthermore, off-central collisions both at the LHC and RHIC, as well as 
lower-energy collisions as those explored at the RHIC energy scan, produce a QGP with a smaller initial temperature. Similarly
 the apparent success of hydrodynamics in smaller systems such as p-Pb \cite{Bozek:2011if} and p-p \cite{Schenke:2014zha,Habich:2015rtj} collisions indicate the need to study the properties of deconfined but cooler QCD plasma, where non-conformal effects become significant (see \cite{Jeon:2015dfa} and references therein for a recent review on the hydrodynamic modelling of heavy ion collisions). 
 
In order to study non-conformal theories in a holographic setup we will consider a five-dimensional bottom-up  model that nevertheless shares many qualitative features with top-down string models. Specifically, our model is dual to a four-dimensional gauge theory that, at zero temperature, flows from an ultraviolet (UV) fixed point to an infrared (IR) fixed point. This renormalisation group (RG) flow is dual on the gravity side to a domain-wall geometry that interpolates between two AdS spaces. The reason why we require that the flow approaches a fixed point in the UV is that this is the situation in which the holographic duality is best understood. The reason for the IR fixed point is that this guarantees that the zero-temperature solution is smooth in the deep IR. The flow is triggered by a source $\Lambda$ for a relevant, dimension-three operator in the UV. We will see that this simple model exhibits a rich phenomenology. In particular, we will study the relaxation of small-amplitude, homogeneous perturbations by computing the spectrum of quasi-normal modes (QNM) with zero spatial momentum. We will see that 
the dominant channel for relaxation in this approximation  depends on the value of the ratio $T/\Lambda$, with $T$ the temperature of the system. At small $T/\Lambda$ the system first EoSizes and subsequently 
isotropises.  In contrast, at large $T/\Lambda$ the order in which these two processes take place is reversed. Although our calculation is done at zero spatial momentum we will argue that, actually, the ordering above is still valid for long-wave-length fluctuations with $k \ll T$.

Previous analyses addressing the near-equilibrium properties of strongly coupled non-abelian plasmas include \cite{Heller:2013oxa,Janik:2015iry,Buchel:2015saa,Janik:2015waa,Buchel:2015ofa,Rougemont:2015wca,Gursoy:2015nza,Gursoy:2016tgf,Ali-Akbari:2016sms}. In particular, the last reference in this list appeared while this paper was being typeset and has some overlap with our observations concerning the different relaxation channels.

%Our analysis is not the first addressing the near-equilibrium properties of strongly coupled non-abelian plasmas. The relaxation of different top-down holographic construction was first studied in \cite{Buchel:2015ofa}. Almost simultaneously, the relaxation of a bottom-up constructions, designed to mimic the QCD equation of state \cite{Gubser:2008ny} was studied in \cite{Janik:2015waa}. Different aspects of this relaxation process was later studied in \cite{Gursoy:2015nza,Gursoy:2016tgf,Rougemont:2015wca,Buchel:2015saa}. In this work we will a bottom-up approach and study a simple holographic construction which may be view as a non-trivial renormalisation  flow from a strongly coupled non-abelian gauge theory in the ultraviolet to a different conformal theory (? with a different central charge?) in the infrared. The flow is generated by a dimension 3 operator in the UV. Although the holographic model is simple, as we will discuss, it exhibits a rich phenomenology. In particular, we will observe that the route for relaxation of this models at high and low temperatures (compared to the source of the operator) differs.

This paper is organised as follows. In Sec.~\ref{sec:model} we introduce the holographic model and discuss its vacuum properties. In Sec.~\ref{sec:thermo} we study black brane solutions  and extract from them the equation of state and the viscosities of the model. In Sec.~\ref{sec:QNM} we study the relaxation of small excitations of the system by computing the QNM spectrum of the black branes at different temperatures and zero spatial momentum.  Finally, in  Sec.~\ref{sec:conclussions} we discuss our main findings and place them in the context of the hydrodynamisation of non-abelian plasmas.

\section{A non-conformal holographic model}
\label{sec:model}
The holographic model that we will consider consists of  five-dimensional Einstein gravity coupled to a scalar field with a non-trivial potential:
\be
\label{eq:action}
S=\frac{2}{\kappa_5^2} \int d^5 x \sqrt{-g} \left[ \frac{1}{4} \R  - \frac{1}{2} \left( \nabla \phi \right) ^2 - V(\phi) \right ] \, ,
%S=\frac{N^2}{2 \pi^2 L^3} \int d^5 x \sqrt{-g} \left[ \frac{1}{4} \R  - \frac{1}{2} \left( \nabla \phi \right) ^2 - V(\phi) \right ] \, ,
\ee
where $\kappa_5$ is the five-dimensional Newton constant.  For specific forms of $V(\phi)$, this action may be viewed as a consistent truncation of five-dimensional  $\mathcal{N}=8$ supergravity. In this paper we will consider a bottom-up model by choosing a potential that is particularly simple and yet shares some of the qualitative properties of these top-down potentials. In particular, we will choose $V(\phi)$ to be negative and to possess a maximum at $\phi=0$ and a minimum at $\phi=\phi_M > 0$. Each of these extrema yields an AdS solution of the equations of motion with constant $\phi$ and radius $L^2=-3/V$. In the gauge theory each of these solutions is dual to a fixed point of the RG with a number of degrees of freedom $N^2$ proportional to $L^3/\kappa_5^2$.\footnote{In the case of $\mathcal{N}=4$ SYM the precise relation would be $L^3/\kappa_5^2=N^2/4\pi^2$.} 

We will be interested in domain-wall solutions interpolating between these two AdS solutions. In the gauge theory, these are dual to RG flows from the UV fixed point at $\phi=0$ to the IR fixed point at $\phi=\phi_M$. 
% For specific form of $V(\phi)$,this action may be viewed as a consistent truncation of five diminutional $N=8$ supergravity. However, in this paper, we will choose an arbitrary  $V(\phi)$ as a way to generate a bottom-up holographic model.. Choosing potentials such that $V(0)=-3/L^2$, the solution to the classical equations of motion associated to the action \eq{eq:action} with trivial $\phi$ is $AdS_5$ with (negative) cosmological constant $\Lambda=-12/L$. With this requirement,  domain-wall solutions  of the classical equations of motion with non-trivial dilation are always asymptotically $AdS_5$. 
 The problem of finding those solutions is significantly simplified if the potential can be written globally in terms of a superpotential, $W$, as
\be
\label{eq:defW}
V(\phi)= -\frac{4}{3} W\left(\phi \right)^2 + \frac{1}{2} W'\left(\phi\right)^2 \,. 
\ee
In this case, vacuum solutions to the Einstein equations can be easily found. Parametrizing the metric as 
\be
\label{eq:dsvac}
ds^2 = e^{2 A(r)} \left(-d t^2 + d{\bf x}^2\right) + dr^2 \, ,
\ee
the solution of the back-reacted gravitational problem is reduced to the first-order equations \cite{Bianchi:2001kw}
\be
\label{eq:fos}
\frac{dA}{dr}  = -\frac{2}{3} W\, , \quad \quad \frac{d \phi}{ d r} = \frac{d W}{ d\phi}\,.
\ee 
We will choose a simple superpotential characterised by a single parameter, $\phi_M$,
\be
\label{eq:defWs}
L\, W\left(\phi \right)=-\frac{3}{2} - \frac{\phi^2}{2} + \frac{\phi^4}{4 \phi^2_M} \,,
\ee
which together with \eq{eq:defW} yields the potential
\be
\label{eq:pot}
L^2 V=-3 -\frac{3}{2} \phi^2 - \frac{1}{3} \phi^4 + \left( \frac{1}{3 \phi_M^2} +  \frac{1}{2 \phi_M^4}\right) \phi^6-\frac{1}{12 \phi_M^4} \phi^8 \,.
\ee
Note that both the superpotential and the potential have a maximum at $\phi=0$ and a minimum at $\phi=\phi_M$. 
This choice leads to three important properties of the associated vacuum solution. First, the resulting geometry is asymptotically AdS$_5$ in the UV with radius $L$,  since $V(0)=-3/L^2$. Second, the second derivative
of the potential at $\phi=0$  implies that, in this asymptotic region, the scalar field has mass $m^2=-3/ L^2$. Following the standard quantisation analysis, this means  that, in the UV,  this field is dual to an operator in the gauge theory, $\O$, with dimension $\Delta_\textrm{UV}=3$. Third, the solution near $\phi = \phi_M$ is again 
AdS$_5$ with a different radius 
\be
\label{eq:LIR}
L_{\rm IR}= \sqrt{- \frac{3}{V\left(\phi_M\right)}} = \frac{1}{1+ \frac{1}{6} \phi_M^2} L \, .
\ee
In this region the effective mass of the scalar field differs from its UV value and it is given by 
\be
\label{eq:MIR}
\mIR= \frac{12}{ L ^2} \left(1+\frac{1}{9} \phi_M^2 \right)= \frac{12}{L^2_{IR}} \frac{\left(1+\frac{1}{9} \phi_M^2 \right)}{\left(1+\frac{1}{6} \phi_M^2 \right)^2}\,. 
\ee
As a consequence, the operator $\O$ at the IR fixed point has dimension 
\be
\label{IRdim}
\Delta_\textrm{IR}=2 + 2\sqrt{ 1+ \frac{\mIR \LIR^2}{4}}=
6\, \left( 1+\frac{\phiM^2}{9}\right) \left(1+\frac{\phiM^2}{6} \right)^{-1}\,.
\ee
To summarize, the vacuum solution describes a RG flow from an UV to an IR fixed point with a smaller number of degrees of freedom, as indicated by the fact that $L_{\rm IR} < L$. We see that changing $\phi_M$ has two main effects. First, as $\phi_M$ increases the difference in degrees of freedom between the UV and the IR fixed points increases. Second, the dimension of the scalar operator at the IR fixed point decreases with increasing $\phi_M$, reaching the marginal dimension $\Delta_\textrm{IR} = 4$ at $\phi_M\to \infty$. However, in this limiting case the IR fixed point disappears and the background solution becomes singular, as is evident from the fact that the effective AdS radius goes to zero as $\phi_M\to \infty$.

Our simple choice of  the superpotential allows us to determine analytically the vacuum solution for arbitrary $\phi_M$. Solving \eq{eq:fos}, we obtain 
\be
\label{eq:metricsol}
e^{2 A}&=& \frac{\Lambda^2 L^2}{\phi^2} \,
  \left(1- \frac{\phi ^2}{\phi _M^2} \right)^{\frac{\phi_M^2}{6}+1} \, 
  e^{-\frac{\phi ^2}{6}}  \,,
\\[2mm]
\label{eq:phisol}
\phi(r)&=& \frac{\Lambda L \, e^{-r/L}}{\sqrt{1+ \frac{\Lambda^2 L^2}{\phi_M^2}e^{-2 r/L} }} \,,
\ee
where $\Lambda$ is an arbitrary constant that controls the magnitude of the non-normalizable mode of the scalar field. As we will see, in the dual gauge theory side, $\Lambda$ is identified with the source of the dimension-3 operator $\O$. The presence of this source  breaks conformal invariance explicitly. 
%Therefore, the vacuum of the theory develops non-vanishing expectation values for the some of the gauge theory operators, such as $\O$ or the stress tensor $T^{\mu \nu}$.  

Noticing that the small field behaviour of the superpotential \eq{eq:defWs} is identical to that of the GPPZ flow \cite{Girardello:1998pd}, we can readily determine the vacuum expectation values (VEV) of the stress tensor and the scalar operator. We begin by  expanding the metric and the scalar field in powers of  $u=Le^{-r/L}$ in the $u \rightarrow 0$  limit. Following \cite{Bianchi:2001kw}, we write the  5-dimensional metric in the form  
\be
ds^2 = \frac{L^2}{u^2} \left (d u^2 + g_{\mu \nu} \, dx^\mu dx^\nu\right) \, ,
\ee
and we write  the power expansion coefficients of the metric and the scalar field as 
\be
\label{eq:pwg}
g_{\mu \nu}&=& \eta_{\mu \nu} + g^{(2)}_{\mu \nu} \, u^2 +  g^{(4)}_{\mu \nu} \, u^4 + ... \, ,
\\[2mm]
\label{eq:pwphi}
\phi &=& \Lambda u \left(1 + \phi_2 u^2 + \ldots\right) \,.
\ee
The expectation values of the field theory operators are then given by %\dm{Change sign of O?}
\be
\label{eq:Texp}
% \left< T_{\mu \nu} \right> &=& 
 %                                                        \frac{2}{\kappa_5^2 \, L} \left(g^{(4)}_{\mu \nu}  + \left(\phi^2_0 \,\phi_2 - C\right)\eta_{\mu \nu}\right) \, , \\
 \left< T_{\mu \nu} \right> &=& 
                                                         \frac{2 L^3}{\kappa_5^2} 
                                                         \left[ g^{(4)}_{\mu \nu}  + \left(\Lambda^2 \,\phi_2 -  \frac{\Lambda^4}{18} + \frac{\Lambda^4}{4\phiM^2}\right) \eta_{\mu \nu}\right] \, , \\[2mm]
\label{eq:Oexp}
%\left< \O \right> &=&  \frac{4}{\kappa_5^2}\,  \phi_0 \phi_2 \, ,
\left< \O \right> &=& - \frac{2 L^3}{\kappa_5^2}\,  \left(2 \Lambda \phi_2  + \frac{\Lambda^3}{\phiM^2} \right)\,.
\ee
To arrive at these expressions we have chosen the superpotential as a counterterm to regularise the on-shell action, which is possible because in our model the superpotential corresponds to a deformation of the gauge theory as opposed to a VEV  \cite{Papadimitriou:2004rz}. We emphasize that these expressions are valid even  if the metric $g_{\mu \nu}$ does not posses the full Poincar\'e symmetry but only rotational and translational invariance along the gauge theory directions, as will be the case for the black brane geometries that we will study in the 
next section. As expected, Eqs.~\eqq{eq:Texp} and \eqq{eq:Oexp} imply the  Ward identity for the trace of the stress tensor  
\be
\label{eq:TTrace0}
\left<T^{\mu}_\mu\right>= - \Lambda \left< \O \right> \, .
\ee

Eqs.~\eqq{eq:metricsol} and \eqq{eq:phisol} determine the VEVs in the vacuum of the theory.  Let us define the energy density $\epsilon$ and the pressure $p$
as the diagonal components of the expectation value of the stress tensor, 
$\left< T^{\mu \nu}\right> ={\rm Diagonal } \left\{\epsilon, p, p, p\right\}$.  The near boundary behaviour of $\phi$, \eq{eq:phisol}, 
leads to 
\be
\phi_2&=& - \frac{\Lambda^2}{2 \phi_M^2} \, ,
\ee
which implies that in the vacuum 
\be
\label{eq:vacvevs}
\left< \O\right> &=&0 , \quad \quad \left< T^{\mu\nu}\right> = 0 \, .
\ee
Note that the explicit breaking of scale invariance means that the trace of the stress tensor is non-zero as an operator.
However, the VEV of this operator vanishes in the vacuum state, as implied by trace Ward identity (2.17) together with the fact
that $\left< \O\right> =0$ in the vacuum for our choice of renormalisation scheme. It should be emphasized that even though the trace Ward identity \eqq{eq:TTrace0}
is scheme-independent, the individual vacuum expectation values of the trace of the stress tensor and of the scalar operator do depend on the renormalisation scheme. In the model we study here the only scheme ambiguity corresponds to a term of the form $\Lambda^4 \, \eta_{\mu\nu}$ in the expectation value of the stress tensor, accompanied by a term of the form $\Lambda^3$ in the expectation value of 
$\O$, with the relative coefficient such that \eqq{eq:TTrace0} is preserved.

\begin{figure}
\begin{tabular}{cc}
\put(40,50){$	\phiM=1$}
\includegraphics[width=.49\textwidth]{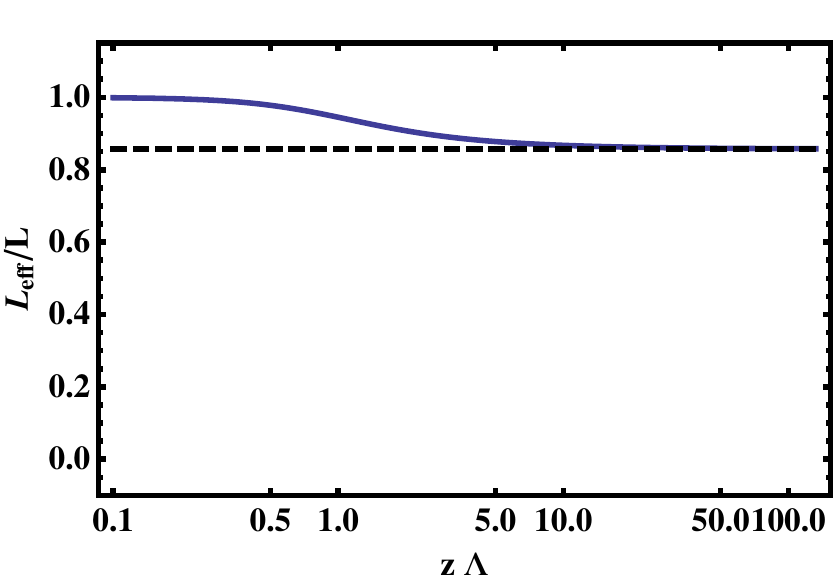} &
\put(40,50){$	\phiM=3$}
\includegraphics[width=.49\textwidth]{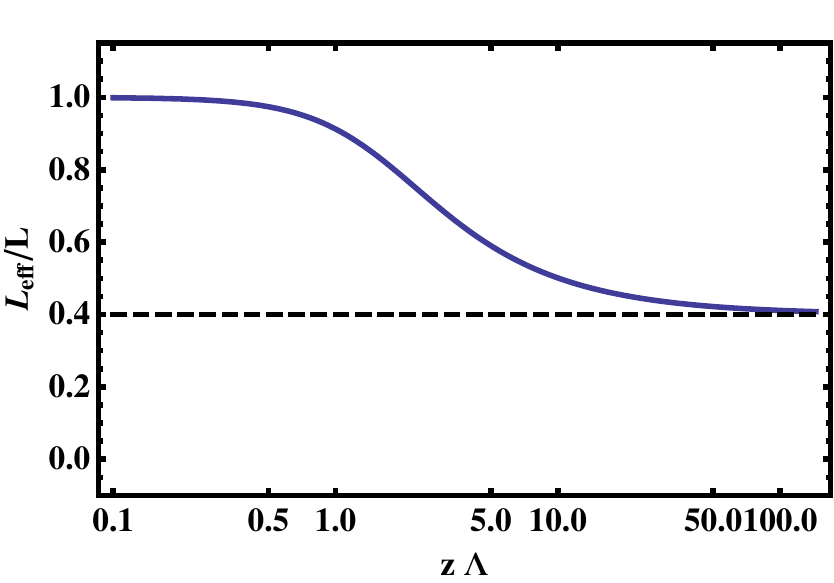}
\\
\put(40,50){$	\phiM=10$}
\includegraphics[width=.49\textwidth]{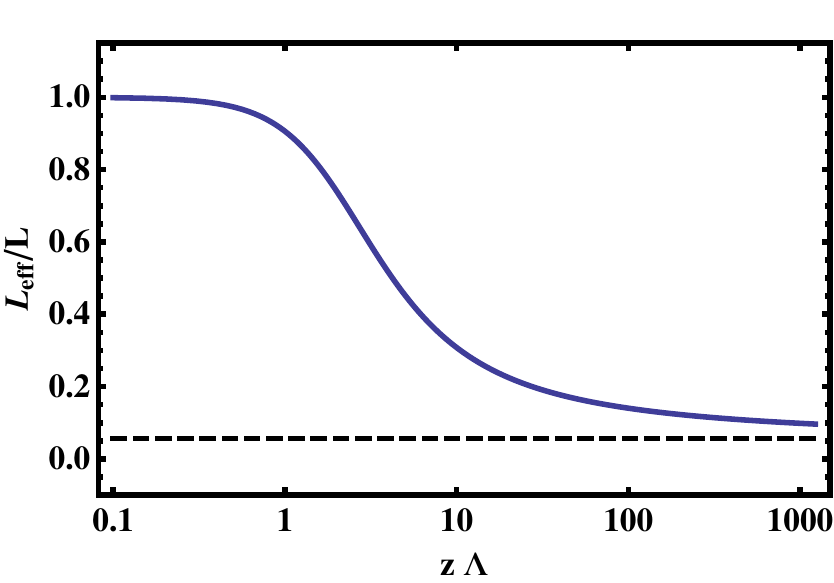} &
\put(40,50){$	\phiM=100$}
\includegraphics[width=.49\textwidth]{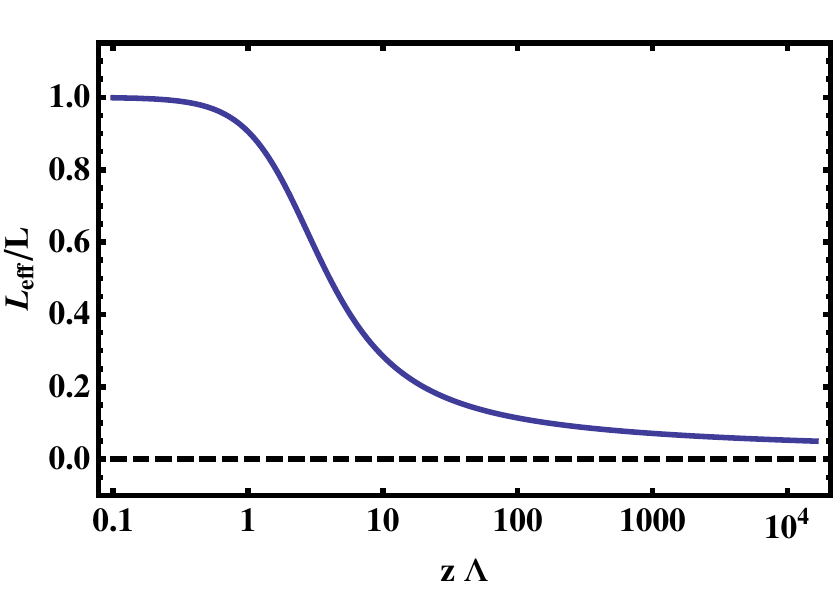}
\end{tabular}
\caption{\label{plot:Leff} $L_{\rm eff}/L$ as a function for $\z$ for different values of $\phiM$. In all panels, the dashed line shows the asymptotic infrared value of the effective AdS radius $L_{\rm IR}$. Note the different scales of the horizontal axes in the different panels.}
\end{figure}

To estimate at which scale non-conformal effects become important, let us perform a change of variables in the holographic direction, which explicitly exploits the relation 
between the dynamics in the bulk with the physics at different scales in the field theory. Denoting the coordinate by $\z$, we write the metric as 
\be
ds^2= \frac{\Leff(\z)^2}{\z^2} \left(-dt^2 + d{\bf x}^2 + d\z^2\right) \,,
\ee
with $\Leff$ a non-trivial function of $z$ such that $\Leff(0)=L$ and $\Leff(\infty)=L_{\rm IR}$. In this set of coordinates, at least in the two asymptotic conformal regions, the coordinate $z$ is related to the energy scale, $Q$, in the gauge theory through $z\sim 1/Q$. The relation between $z$ and $u$  is given by 
\be
\z(u)= \int_0^u  du \frac{L}{u} e^{-A}\, ,
\ee
and the function $\Leff$ is given by
\be
\Leff (\z)= \z \,e^{A} \,.
\ee
In Fig.~\ref{plot:Leff}  we show the ratio $L_{\rm eff}/L$ as a function of $\z$  for several different values of the  parameter $\phi_M$ controlling the physics of the model. We see that the system behaves approximately conformally up to scales of order $z \sim \Lambda$. At this scale, the metric starts to deviate significantly from that of $AdS_5$, and $L_{\rm eff}$ decreases as a function of $\z$. Sufficiently deep in the IR, $L_{\rm eff}$ approaches $L_{\rm IR}$ and the system behaves again as approximately conformal. However, the scale at which this transition occurs depends significantly on the model parameter $\phi_M$; as $\phi_M$ increases, the function $L_{\rm eff}$ approaches its asymptotic value more slowly. 
This different rates at which the IR fixed point is approached have consequences for the finite-temperature behaviour of the dual gauge theory, as we will see in the next section.

\section{Thermodynamics and transport}
\label{sec:thermo}
We will now explore the thermal physics of the gauge theory dual to the gravitational model described in the previous section.\footnote{Previous studies of the thermodynamics of Einstein+scalar gravity include \cite{Megias:2010ku,Veschgini:2010ws,Gursoy:2008za}.} To do so, we will search for black brane solutions of the action \eqq{eq:action}. We will follow the method of the master function, introduced in ref.~\cite{Gubser:2008ny}, to which we refer the reader for details.\footnote{Note that our normalisations of the scalar field and of the potential differ from those in \cite{Gubser:2008ny}.}
Since for the background solution \eqq{eq:phisol} the scalar field is a monotonic function of $u$, we may use the scalar field as a coordinate and express the metric as 
\be
\label{eq:thermal_line_element}
ds^2= e^{2 A}\left(-h(\phi) d\tau^2 + d{\bf x}^2 \right) -2 e^{A+B} L \, d\tau d\phi \, ,
\ee
with $h(\phi)$ vanishing at $\phi=\phiH$, the value of the scalar field at the horizon, i.e.~$h(\phiH)=0$. 
The region outside the horizon corresponds to 
$0<\phi< \phiH$.
For later convenience, we have expressed the metric in Eddington-Finkelstein form. 
With this ansatz,  Einstein's equations take the form 
\be
\label{eq:Einstein}
A''(\phi )-A'(\phi ) B'(\phi )+\frac{2}{3}&=&0  \,, \nonumber \\
4 A'(\phi ) h'(\phi )-B'(\phi ) h'(\phi )+h''(\phi
   )&=&0 \,,\\
 \frac{3}{2} A'(\phi ) h'(\phi )+h(\phi ) \left(6
   A'(\phi )^2-1\right)+2 e^{2 B(\phi )}
 L^2 V(\phi )&=&0  \,,
\nonumber \\
4 A'(\phi )-B'(\phi )-\frac{e^{2 B(\phi )} L^2
  V'(\phi )}{h(\phi )}+\frac{h'(\phi
   )}{h(\phi )}&=&0     \,. \nonumber 
\ee 
 A solution to these  equations may be found in terms of a master function $G(\phi)$ defined as 
\be
G(\phi)= \frac{d}{d\phi} A(\phi) \, .
\ee
Manipulating the set of equations \eqq{eq:Einstein}, a non-linear equation for $G$ was found in \cite{Gubser:2008ny}:
\be
\label{eq:master}
\frac{G'(\phi )}{G(\phi )+\frac{4 V(\phi )}{3 V'(\phi )}}
=
\frac{d}{d\phi} \log \left(
\frac{1}{3 G(\phi )}-2 G(\phi )+\frac{G'(\phi )}{2 G(\phi )}-\frac{G'(\phi )}{2 \left(G(\phi )+\frac{4 V(\phi )}{3 V'(\phi )}\right)} 
\right) \, .
\ee
Close to the boundary, $\phi\rightarrow0$, the solution of this equation behaves as 
\be
G(\phi)= \frac{1}{\Delta -4} \frac{1}{\phi} + \cdots \,,
\ee
with $\Delta$ the scaling dimension of the dual operator. With our choice of potential \eqq{eq:pot} we have  $\Delta=3$. 
Using  \eq{eq:Einstein}, the different metric coefficients are given by 
\be
\label{eq:A}
A(\phi)&=&-\log \left(\frac{\phi}{\Lambda}\right) + \int_0^\phi d\phit \left( G(\phit) + \frac{1}{\phit }\right) \, ,
\\[2mm]
\label{eq:B}
B(\phi)&=& \log \left( \left| G(\phi) \right|\right) + \int_0^\phi d\phit \frac{2}{3 G(\phit)} \, ,\\[2mm]
\label{eq:h}
h(\phi)&=&-\frac{e^{2 B(\phi )} L^2 \left(4V(\phi )+3 G(\phi )V'(\phi )\right)}{3 G'(\phi )} \, .
\ee
In these expressions, the constants of integration are fixed by requiring that, close to the boundary, the metric and scalar field may be expressed as in \eq{eq:pwg} and \eq{eq:pwphi} .
At the horizon, the condition $h(\phiH)=0$ together with the last two equations in \eqq{eq:Einstein} fix the value of $G(\phiH)$. Starting from this fixed value, a power series solution close to the horizon may be found as
\be
G(\phi)=
   -\frac{4
   V(\phiH)}{3 V'(\phiH)} + 
 \frac{2}{3} ( \phi - \phiH)
   \left(\frac{V(\phiH) V''(\phiH)}{V'(\phiH)^2}-1\right) + \mathcal{O} \left(\left(\phi-\phiH\right)^2\right) \,.
\ee

From these metric coefficients, we can extract the Hawking temperature $T$ and the entropy density $s$ of the black brane:
\be
LT= \frac{A(\phiH )-B(\phiH)}{4 \pi}\,,   \quad \quad  s= \frac{2 \pi}{\kappa_5^2} e^{3 A(\phiH)} \,.
\ee
The relation of the different metric coefficients with the master function leads to the following form for the temperature and entropy of the thermal state:
\be
%T&=&-\frac{L V(\phiH)}{3 \pi} \exp \left\{A (\phiH) + \int_0^\phiH d\phi \frac{2}{3 G(\phi)} \right\}
%\\
T&=&-\Lambda \frac{L^2 V(\phiH)}{3 \pi \phiH} \exp \left\{ \int_0^\phiH d\phi \left( G(\phi) + \frac{1}{\phi} +\frac{2}{3 G(\phi)}  \right)\right\} \, ,
\\[2mm]
s&=& \frac{2 \pi}{\kappa^2_5}  \frac{\Lambda^3}{ \phiH^{3}} \exp\left\{ 3 \int_0^\phiH d\phi  \left(G(\phi) + \frac{1}{\phi}\right)\right\} \,. 
\ee
These expressions are well suited for the determination of these two quantities from the numerical evaluation of the master equation \eqq{eq:master}. 

\begin{figure}
\includegraphics[width=.49\textwidth]{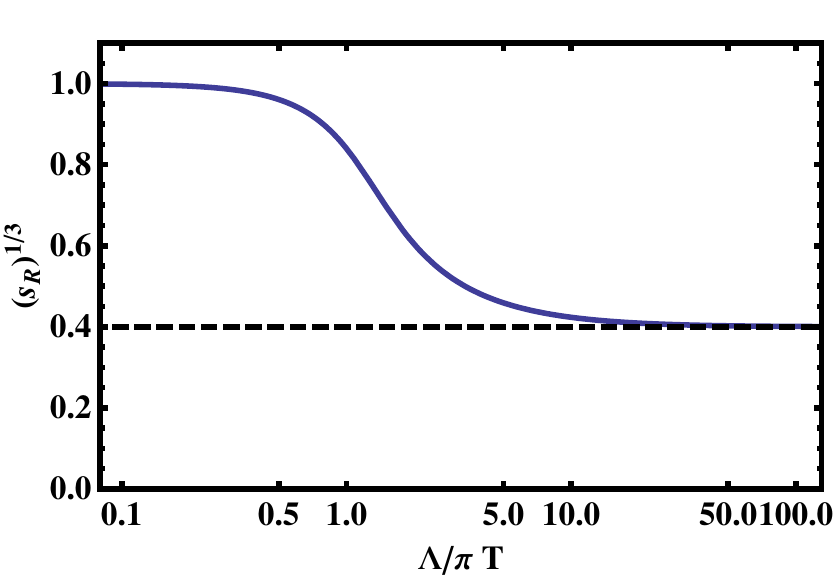}
\includegraphics[width=.49\textwidth]{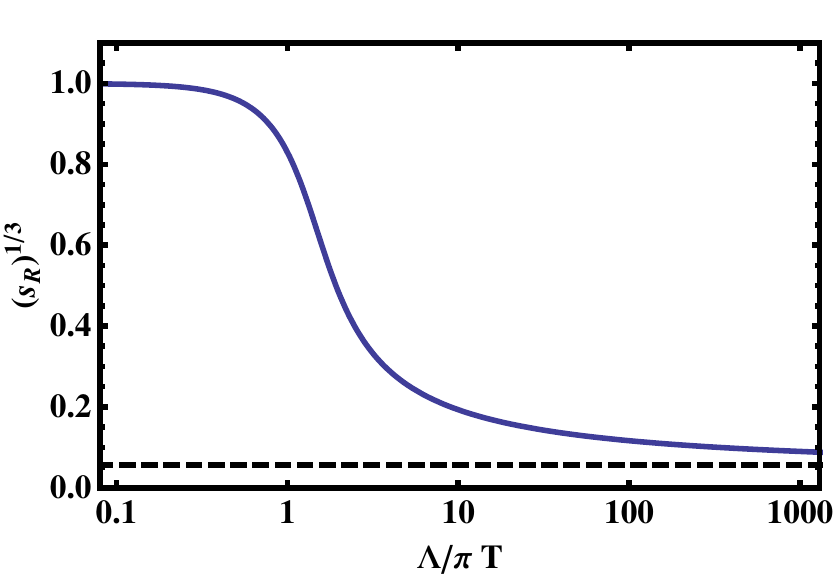}
\caption{\label{plot:sofT} Ratio of entropy density to temperature for   $\phi_M=3$ (left) and $\phi_M=10$ (right) as a function of the inverse temperature. The dashed line shows $L_\mt{IR}/L$. %\dm{Change label on vertical axes to $s_R$.}
} 
\end{figure}

In Fig.~\ref{plot:sofT} we plot the dimensionless quantity 
\be
s_R = \frac{\kappa_5^2}{2 \pi^4 L^3} \frac{s}{T^3}\, , 
\ee
as a function of the inverse temperature for two different values of $\phi_M$. Since the theory is conformal both at the UV and at the
IR, the high and low temperature behaviour of the entropy density must coincide with that of a relativistic conformal
theory and scale as $T^3$. In particular, for a relativistic CFT,  $s / T^3$ is proportional to the number of degrees of freedom in the theory, which for an $SU(N)$ gauge theory with matter in the adjoint representation scales as $N^2$. For example, for $\mathcal{N}=4$ SYM
\begin{equation}
\frac{s}{T^3}= \frac{\pi^2}{2}N^2,
\end{equation}
but the precise coefficient depends on the specific theory. In terms of the parameters of the dual gravity description this quantity becomes
\begin{equation}
\frac{s}{T^3}=\frac{2\pi^4 L^3}{\kappa_5^2}.
\end{equation}
In our bottom-up setup, the above argument  allows us to {\em define} the number of degrees of freedom at the fixed points holographically in terms of the effective AdS radius. In particular, the quantity $s_R$ should approach 1 at high temperature and $(L_\mt{IR}/L)^3$ at low temperature, which is confirmed by the plots in Fig.~\ref{plot:sofT}.

\begin{figure}[t]
\begin{center}
\includegraphics[width=.65\textwidth]{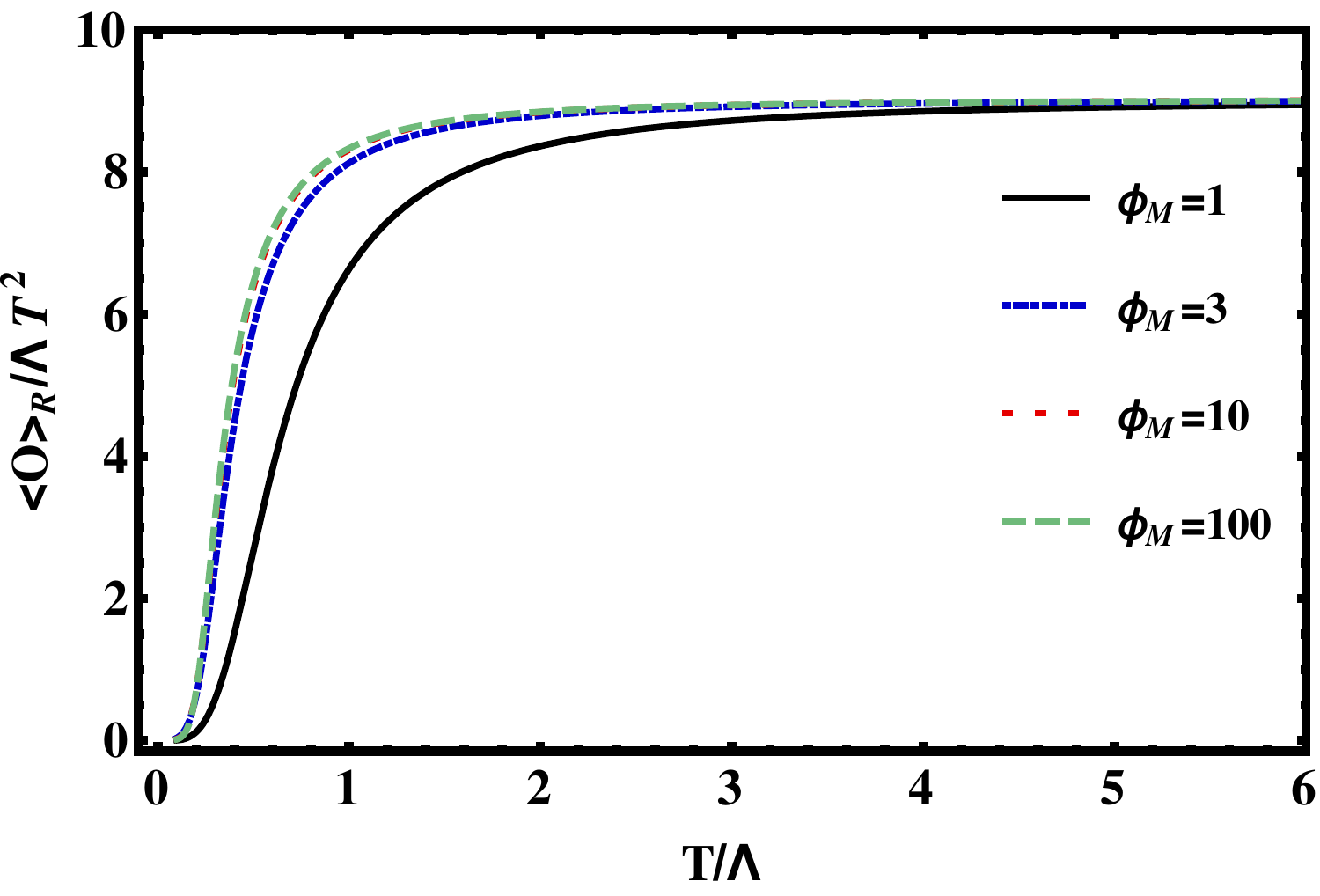}
\caption{\label{plot:phi2} Temperature dependence of the VEV of the scalar operator 
$\left< \O\right>_T$ for several values of $\phi_M$. $\left< \O\right>_R=\kappa_5^2 \left< \O \right>_T /L^3.$ 
%\dm{Define $\left< \O\right>_R$}
}
\end{center}
\end{figure}

Using standard thermodynamic relations and the fact that in our renormalisation scheme  the vacuum pressure is zero 
we can determine the pressure and the energy density of the thermal system through
\be
p= \int_0^T d\tilde T \, s(\tilde T)   \, , \quad \quad \epsilon+p= T s\, .
\ee
Since the theory is not conformal, the trace of the stress tensor in the thermal ensemble does not vanish. Using the Ward identity \eqq{eq:TTrace0}, the   energy density, the pressure and the scalar condensate at non-zero temperature are related through 
\be
\label{eq:TTrace}
\epsilon -3p=  \Lambda \left< \O \right>_T \, .
\ee
The thermal expectation value $\left< \O \right>_{T}$ may be determined from the normalisable mode of the scalar field in the thermal background via \eq{eq:pwphi}.
Since at $T=0$ the scalar VEV vanishes (see \eq{eq:vacvevs}) this relation implies that $\epsilon=3 p$, as expected from the fact that the IR  theory is conformal. 
At  $T>0$, however, $\left< \O \right>_T \neq 0$, as shown in \fig{plot:phi2}, and the expectation value of the trace of the stress tensor does not vanish. Note that, unlike at low temperatures, at which  $\left< \O \right>_T$ depends on $\phiM$,  at high temperatures $\left< \O \right>_T$ becomes independent of $\phiM$. This   is easy to understand from the gravitational computation. At high temperatures the value of the scalar field at the horizon is small and, therefore, the physics is sensitive only to the small-field behaviour of the scalar potential, which is independent of $\phiM$. 
In this limit, the plots in \fig{plot:phi2} show that the VEV scales as $\left< \O \right>_T \sim \Lambda T^2$.
%
% From the field theory side, at temperatures much larger than  the source $\Lambda$, we also expect the theory to become conformal, and $\left< \O \right>_T \sim \Lambda T^2$ (CHECK!!).

\begin{figure}[t]
\begin{center}
\includegraphics[width=.69\textwidth]{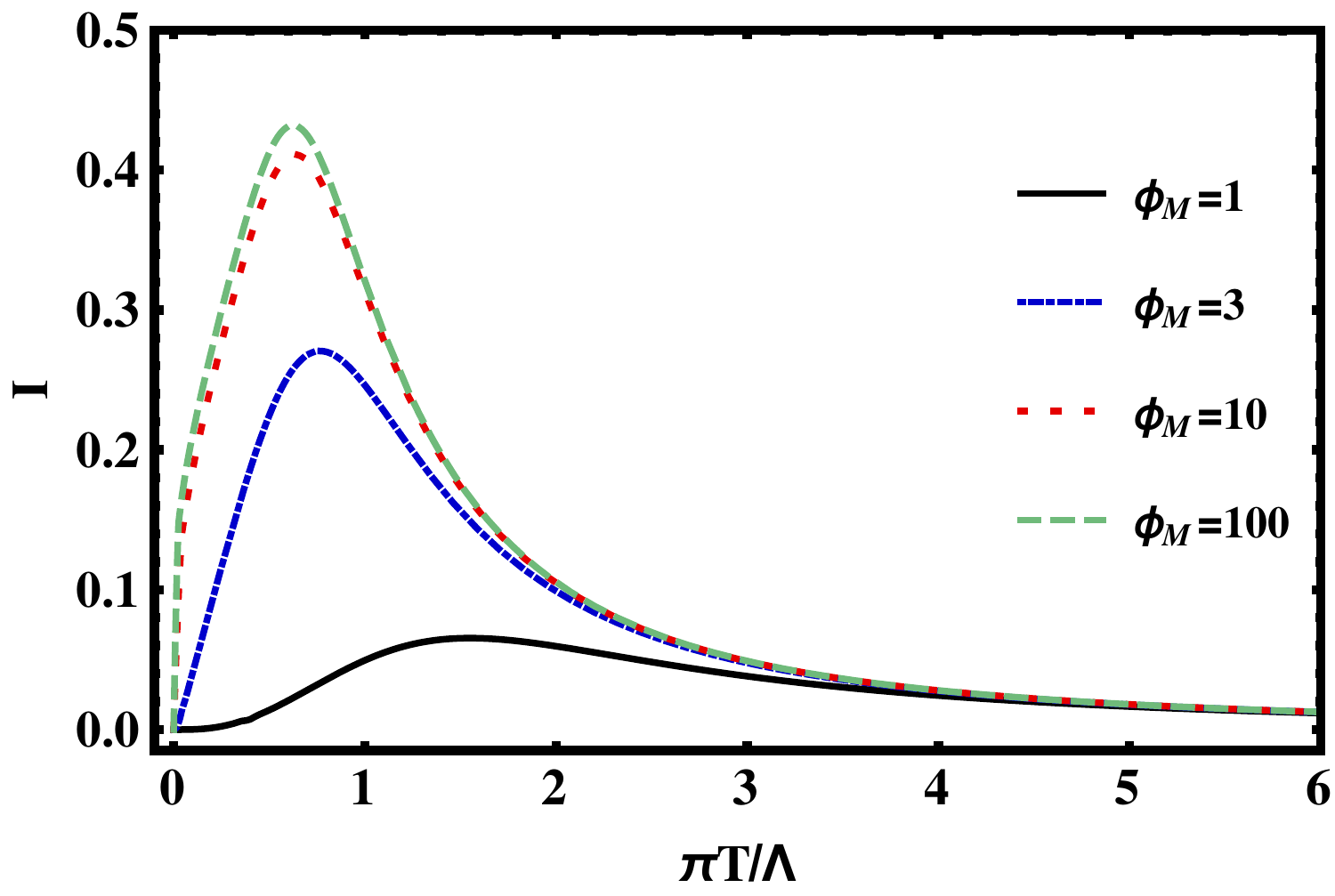} 
\caption{\label{plot:I} Ratio of the trace of the stress tensor to the enthalpy, $I$, as a function of  $T$ for different values of $\phiM$. }
\end{center}
\end{figure}

Despite the fact that the trace of the stress tensor at high temperature does not vanish, the theory does behave as a conformal theory.  From the gauge theory viewpoint this may be understood from the relative 
magnitude of the trace of the stress tensor compared to the energy density or the pressure: while at large $T$ the latter quantities scale as $T^4$, the trace only grows as $T^2$. In \fig{plot:I} we show the temperature dependence of the ratio of the stress tensor to the enthalpy,
\be
\label{eq:Idef}
I=\frac{\epsilon-3p}{\epsilon+p} \,,
\ee
which in the thermal-QCD literature is sometimes referred to as the interaction measure. As anticipated, both at low and high temperatures this ratio vanishes, indicating that the theory becomes effectively conformal in these limits. At intermediate temperatures, the value of $I$ is non-zero and depends on $\phiM$. As inferred from the behaviour of the entropy, the larger $\phiM$ the larger the deviations from conformality in the thermodynamic properties of the theory. Because of this behaviour we may use $I$ as a measure of the non-conformality of the theory. 

\begin{figure}[t]
\begin{center}\includegraphics[width=.69\textwidth]{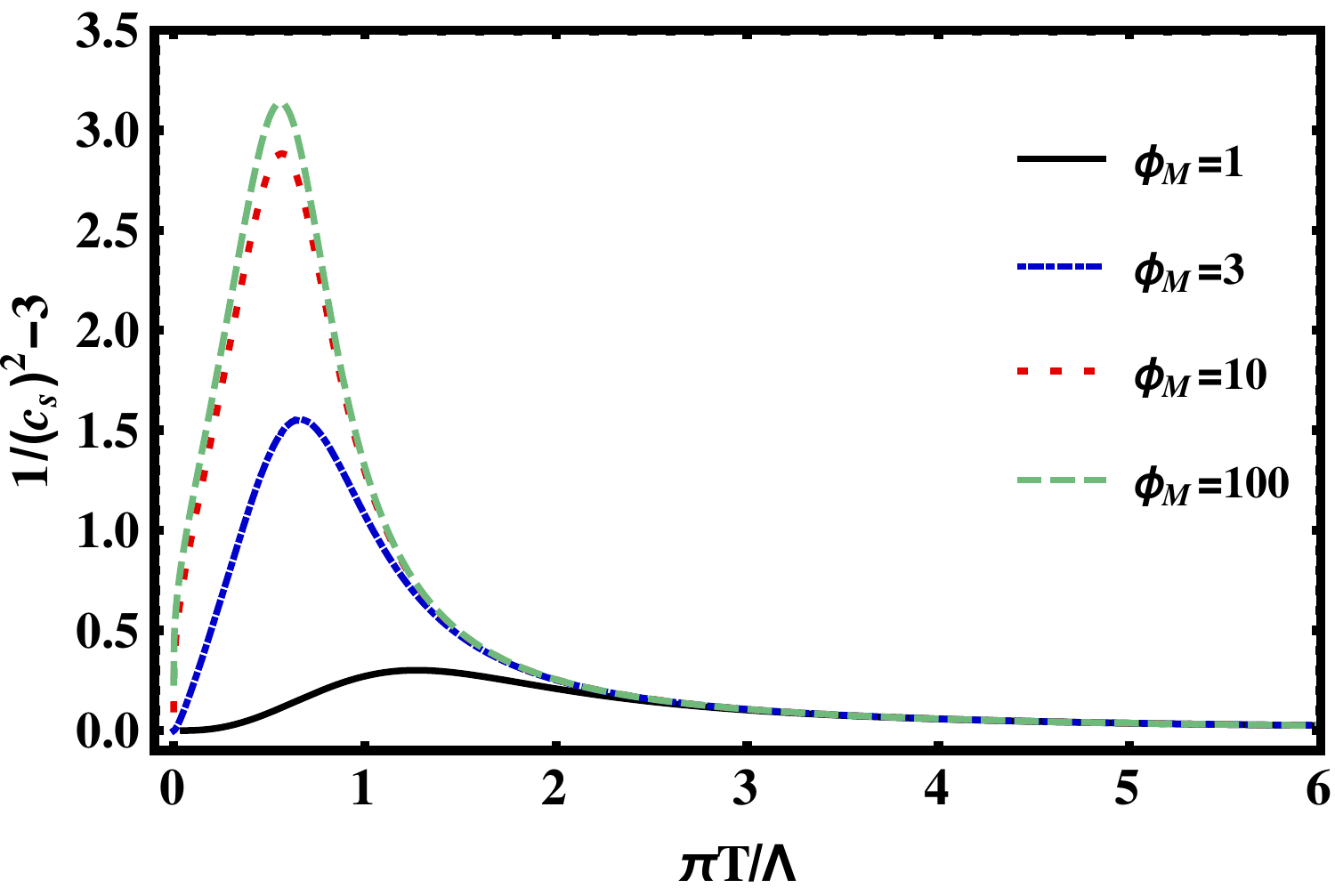} 
\caption{\label{plot:cs} Inverse speed of sound square  as a function of  $T$ for different values of $\phiM$. }
\end{center}
\end{figure}

Another way to quantify the non-conformal behaviour of the thermodynamics of the dual theory is the value of the speed of sound. Using thermodynamic identities, the square of the speed of sound may be determined from the inverse of the logarithmic derivative of the entropy, 
\be
\frac{1}{c_s^2}=\frac{d\log s}{d \log T} \,. 
\ee
In \fig{plot:cs}  we show the temperature behaviour of the deviation of $c_s$ from its conformal value, $c_s=1/\sqrt{3}$, for different values of $\phiM$. The qualitative behaviour of this quantity is very similar to that of $I$. Both at high and low temperatures, the speed of sound approaches its conformal value. At intermediate temperatures we have $c_s^2<1/3$ and the deviation from the conformal value grows with $\phiM$. 

%\begin{figure}
%\begin{tabular}{cc}
%\put(150,50){$	\phiM=1$}
%\includegraphics[width=.49\textwidth]{Plots/cs_phiM1.pdf} &
%\put(150,50){$	\phiM=3$}
%\includegraphics[width=.49\textwidth]{Plots/cs_phiM3.pdf}
%\\
%\put(150,50){$	\phiM=10$}
%\includegraphics[width=.49\textwidth]{Plots/cs_phiM10.pdf} &
%\put(150,50){$	\phiM=100$}
%\end{tabular}
%\caption{\label{plot:cs} $L_{\rm eft}/L$ as a function for $\z$ for different values of $\phiM$. In all panels, the dashed line shows the asymptotic infra-reed value of the effective AdS radius $L_{\rm IR}$. Note the different scale of the x-axis in the different panels}
%\end{figure}

\begin{figure}[t]
\begin{center}
\includegraphics[width=.69\textwidth]{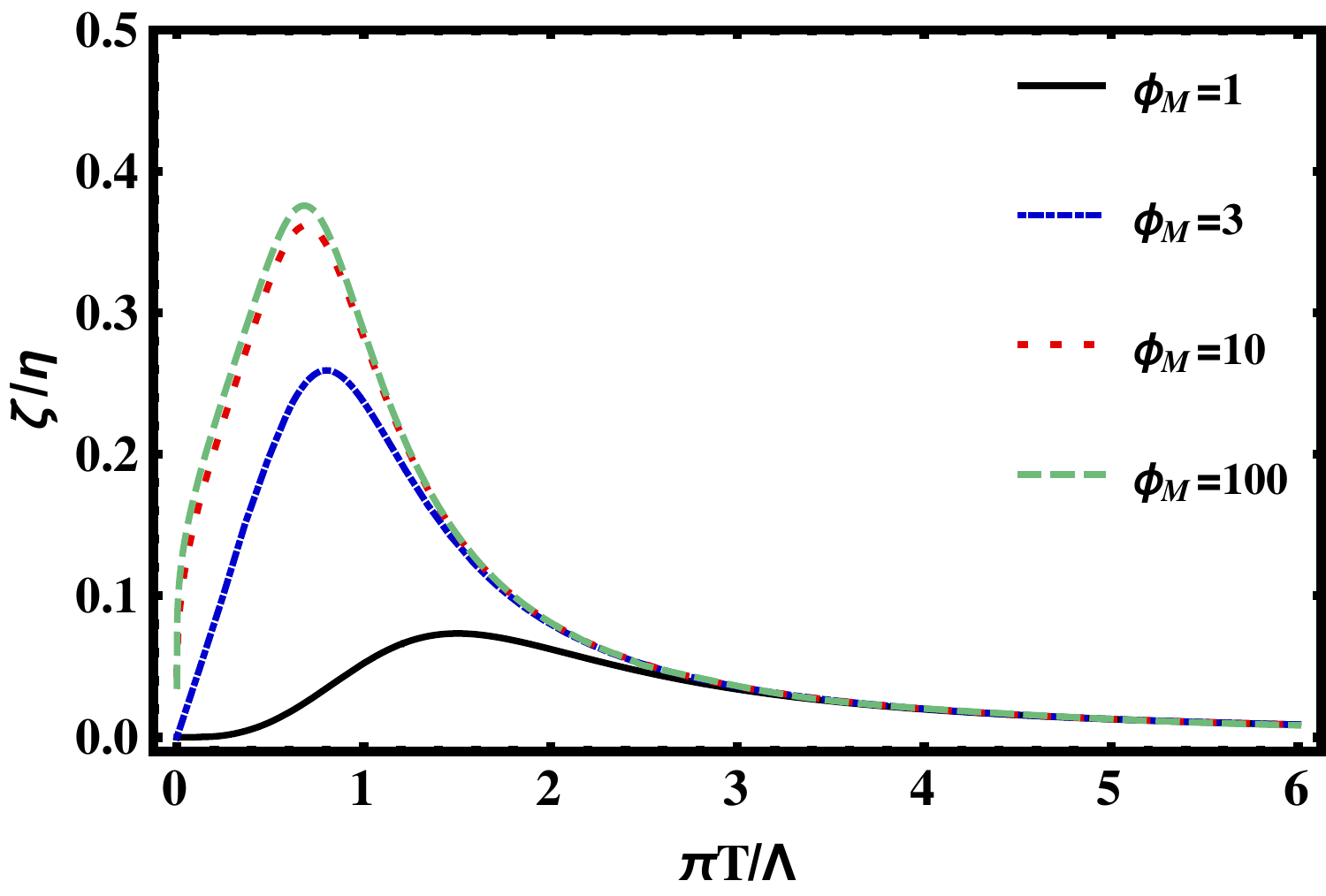}
\caption{\label{plot:zetta}  Ratio of bulk to shear viscosity as a function of temperature for different values of $\phiM$.  }
\end{center}
\end{figure}

The non-conformal behaviour already observed in the equation of state of the system is also reflected in the transport properties of the dual gauge theory plasma.  Since this is isotropic, at leading order in gradients transport phenomena are controlled by only two  coefficients, the shear viscosity $\eta$ and the bulk viscosity $\zeta$. Because of the universality of the shear viscosity to entropy ratio \cite{Kovtun:2004de} in all theories with a two-derivative gravity dual, we have that this ratio in our model takes the same value as in the conformal $\mathcal{N}=4$ theory, i.e.~$\eta/s=1/4\pi$.  In contrast, the bulk viscosity, which would vanish identically in a CFT, is non-zero in our model. 
Following\footnote{Note that our normalisation of the scalar field differs from that in \cite{Eling:2011ms}.} \cite{Eling:2011ms} we determine the bulk viscosity by studying the dependence of the entropy on the value of the scalar field at the horizon,\footnote{We have cross-checked the result of this computation with a two-point function computation as in \cite{Gubser:2008sz}. See also \cite{Mas:2007ng} for a general analysis of the bulk viscosity for Dp-brane solutions.}
\be
\frac{\zeta}{\eta}=4 \left(\frac{d \log s}{d\phiH } \right)^{-2} \, .
\ee
The temperature dependence of this ratio is shown in \fig{plot:zetta} for different values of $\phiM$. The behaviour of this ratio is very similar to that of the interaction measure and the speed of sound:  both at low and high temperatures the ratio of the two viscosities vanishes, while at intermediate temperatures $T \sim \Lambda$ it attains $\phiM$-dependent values that grow with $\phi_M$.  As in the case of $\epsilon -3p$ and the interaction measure, the fact that the ratio of viscosities vanishes at high temperatures does not imply that the bulk viscosity itself vanishes. In fact, we have checked numerically that at high temperatures 
the bulk viscosity scales as $\zeta \sim \Lambda^2 T$. Nevertheless, the fact that the ratio of viscosities approaches zero shows that transport is effectively conformal, since the contribution to the hydrodynamic stress tensor of the bulk tensor is suppressed with respect to the shear one.\footnote{Here we are implicitly assuming that the magnitude of the shear tensor is not parametrically suppressed with respect to the bulk one. Should the flow of the system be prepared such that the shear tensor identically vanishes, then transport would be dominated by the bulk tensor.}

\begin{figure}[t]
\begin{center}
\includegraphics[width=.69\textwidth]{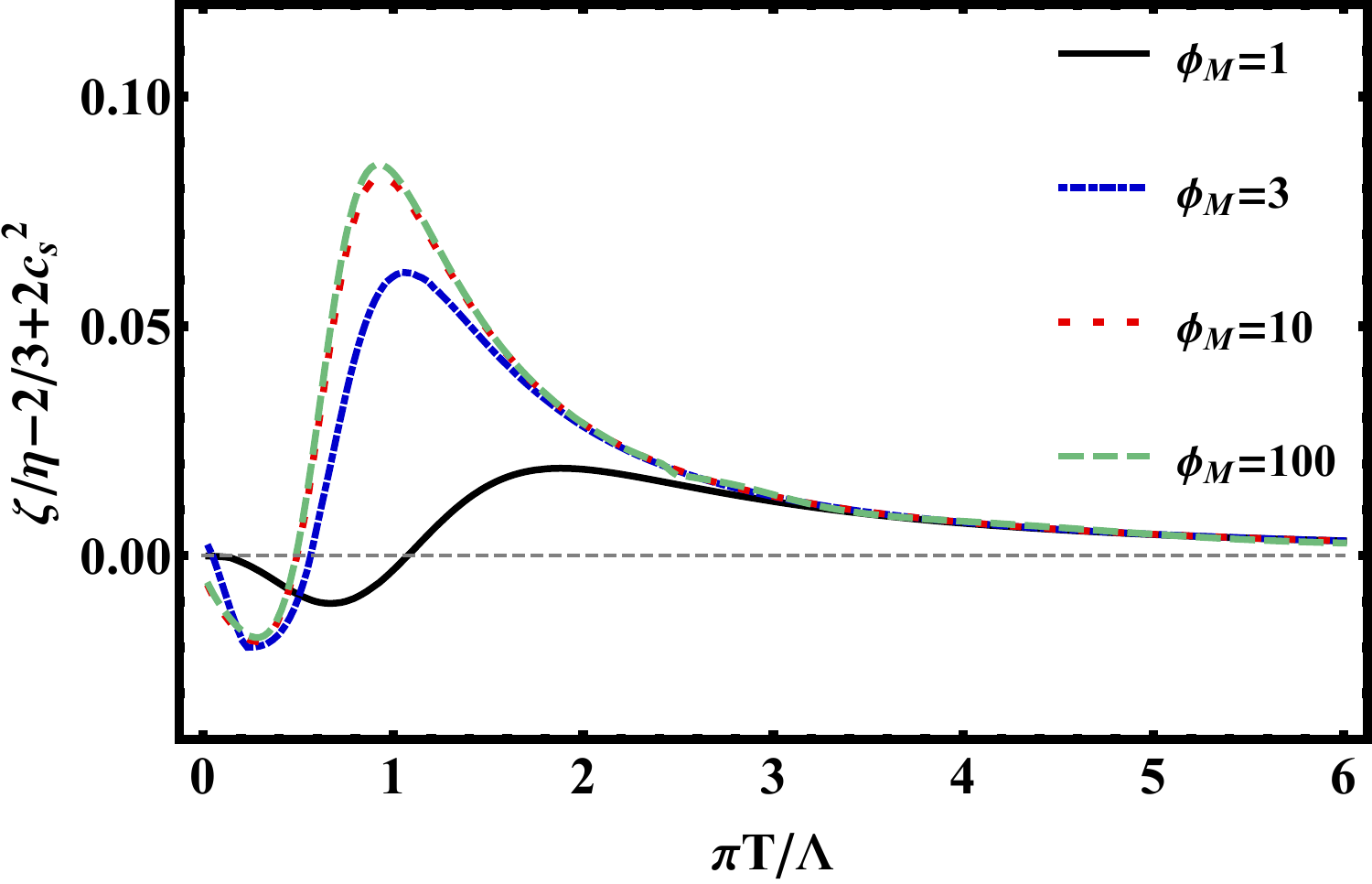}
\caption{\label{buchel}  Violation of Buchel's bound at low temperatures.}
\end{center}
\end{figure}

It is interesting to note that the ratio of viscosities at low temperatures violates  Buchel's bound 
\be
\frac{\zeta}{\eta} \geq 2 \left( \frac{1}{3} - c_s^2 \right) \,,
\ee
as illustrated in \fig{buchel}. Violations of this bound have been previously encountered in other models such as \cite{Gubser:2008ny,Gubser:2008sz,Buchel:2011uj}.

\section{Quasi-normal modes and relaxation}
\label{sec:QNM}
We now turn to the description of the off-equilibrium dynamics of our holographic model. We study the reaction of the system to small perturbations which drive it away from local equilibrium. On the gravity side this problem translates into the study of the 
relaxation of the black brane solutions constructed above when  the different background fields are perturbed. As  is well known, this relaxation process is controlled by an infinite set of discrete, damped modes known as QNMs. In this section we will determine the QNM frequencies of the system as a function of the  temperature. 

%We focus on excitations that change the stress-energy tensor in the dual field theory. In the gauge theory side, these are dual to the dynamics of small fluctuations in the space-time metric. Furthermore, 

Since  in our holographic model the scalar field backreacts on the geometry, metric fluctuations couple to fluctuations of the scalar field and they must all be considered simultaneously. Denoting by $\GT$ the black brane metric in Eddington-Finkelstein coordinates  \eqq{eq:thermal_line_element}, we will study fluctuations of the form
\be
\GT_{M N} \rightarrow \GT_{M N} + h_{M N} \,,  \quad \quad \phi\rightarrow \phi + \varphi \,.
\ee
The dynamics of  $h_{M N}$ and $\varphi$ is governed by the linearised Einstein and scalar field equations on the background spacetime  $\GT_{M N}$. We will use 
the value of the unperturbed scalar field $\phi$ as a coordinate in the holographic direction. 

As is well known (see e.g.~\cite{Kovtun:2005ev}) not all fluctuations are physical, since reparametrisation invariance leads to a gauge symmetry in the linearised equations of motion.  In the presence of a scalar field, the linearised equations of motion are invariant under  the transformation
\be
h_{M N} \rightarrow h_{M N} + \nabla_M \chi_N + \nabla_M \chi_N\, , \quad \quad  \varphi \rightarrow \varphi + \chi^M \nabla_M \phi \,,
\ee
with $\chi^M$ a spacetime-dependent vector field and $\nabla_M$ the covariant derivative in the background metric $\GT_{M N}$. Because of this symmetry, not all fluctuations are physical and the relaxation dynamics of the black brane is encoded in the spectrum of gauge invariant combinations of fields.\footnote{See \cite{Benincasa:2005iv} for a detailed classification of these fluctuations in the context of non-conformal  theories.}

In this paper we will study the relaxation of homogeneous disturbances of the  plasma. In other words, we will allow for time but not for space dependence of the perturbations. We will consider both isotropic and anisotropic perturbations and we will denote by $z$ the direction of anisotropy.  On the gravity side, these  perturbations will depend on time and on the holographic radial coordinate. Under these conditions, there are only 
 two  independent sets of gauge invariant excitations of the plasma, which  may be parametrized by the following combination of fields\footnote{Anisotropic fluctuations induced by $\Zt=h_{xy}$ are also possible and independent of the two modes listed  in \eqq{considered1}-\eqq{considered2}. However, we will not consider these fluctuations here because  at zero spatial momentum the dynamics of $\Zt$ is identical to that of $\Zm$.}
\be
\label{considered1}
\Zm&=&e^{-2 A} \left(h_{zz} - h_{aa}\right) \, ,
\\
\Zphi &=& \varphi -\frac{e^{-2 A(\phi )}}{2 A'(\phi )} h_{aa} \, ,
\label{considered2}
%\\
%\Zt&=&h_{x y} \, ,
\ee
where $h_{aa}=(h_{xx}+h_{yy})/2$.  
The first fluctuation, $\Zm$, controls anisotropic perturbations that leave unaffected the expectation value of the scalar operator, the average pressure and the trace of the stress tensor. The non-conformal mode $\Zphi$ controls fluctuations that change the three pressures in an isotropic way and at the same time modify the expectation value of the scalar operator and the trace of the stress tensor.  At non-zero spatial momentum these excitations would be coupled to one another and they would include the hydrodynamic modes.  Our restriction to the space-independent sector implies that the energy density of the plasma is unchanged by the fluctuations \eqq{considered1}-\eqq{considered2}, since in a homogeneous plasma conservation of the stress tensor reduces to $\partial_t \epsilon =0$.

Manipulating the linearised Einstein and Klein-Gordon equations and after a Fourier transform in time,  the dynamics of the $\Zm$ and $\Zphi$ modes are given by the equations
\be
-i \omega \Llphi \, \Zphi&=& \Lrphi \, \Zphi \, ,
\\
- i \omega \Llm \, \Zm&=& \Lrm  \, \Zm \, ,
\ee
where $\Llphi $, $\Llphi$, $\Lrm$, $\Lrm$ are linear operators in the holographic direction given by 
\be
\Llm&=&\Llphi = \frac{L e^{B-A} }{h(\phi )}  \left(3 A'+2
   \frac{d}{d\phi}\right)\, ,
\\[2mm]
\Lrm&=&
\frac{L^2 e^{2 B} V' }{h} \frac{d}{d\phi}+\frac{d^2}{d\phi^2}  \,, 
\\[2mm]
\Lrphi&=& \Lrm+ 
\frac{\left[8 h \left(6 \left(A'\right)^2-1\right)-3 A' \left(L^2 e^{2 B} \left(3
   A' V''+8 V'\right)-4
   h'\right)\right]}{9 h \left(A'\right)^2} \,,
\ee
with
$A$, $B$ and $h$ the numerically computed functions which determine the background, given by \eqq{eq:A}-\eqq{eq:h}. 
The equation for the anisotropic fluctuation, $\Zm$, is that of a massless probe scalar field, while the equation for the bulk fluctuations $\Zphi$ includes an explicit  dependence on the potential.
 The discrete set of normalizable, in-falling solutions of this system of equations 
are the QNMs.  The fact that the equations are  
linear in the frequency  is a consequence of the Eddington-Finkelstein form of the thermal metric \eq{eq:thermal_line_element}. Following \cite{Janik:2015waa} we use this  to determine the QNMs and their associated frequencies by spectral methods, which allow us to reduce the problem of finding the complex-valued spectrum of excitations to an eigenvalue problem.  This method is particularly suited for background metrics which are only known numerically. We have also double-checked the results for some representative frequencies with a shooting method. 

\begin{figure}
\begin{tabular}{cc}
\put(150,110){$	\phiM=1$}
\includegraphics[width=.49\textwidth]{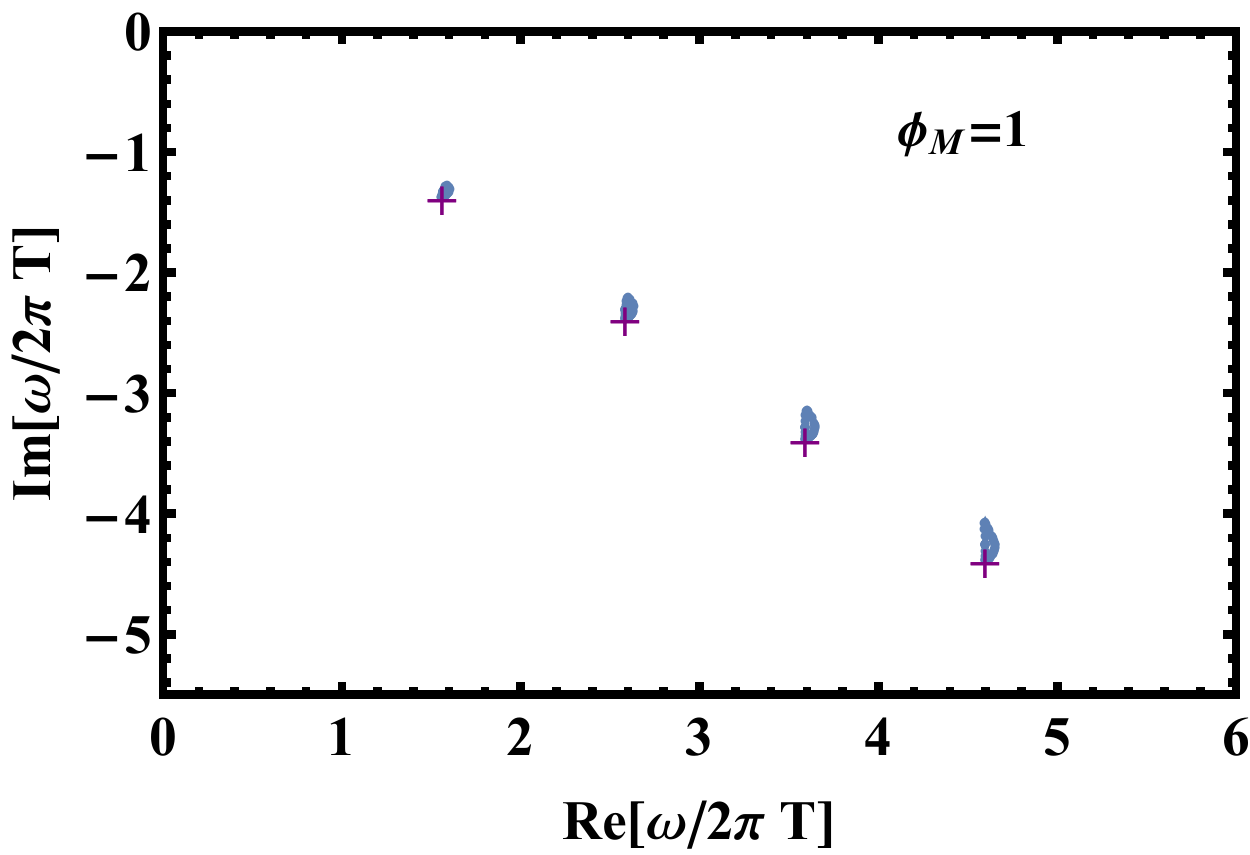} &
\put(150,110){$	\phiM=3$}
\includegraphics[width=.49\textwidth]{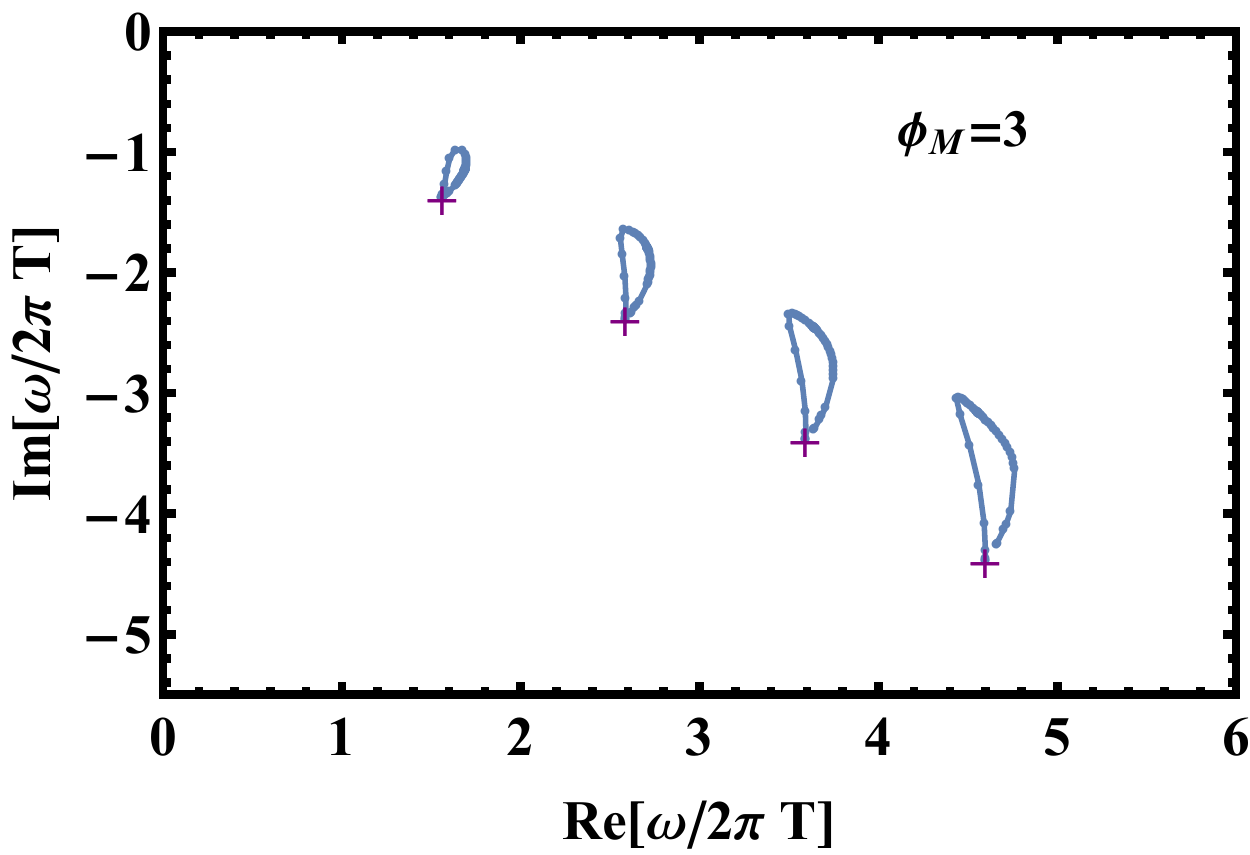}
\\
\put(140,110){$	\phiM=10$}
\includegraphics[width=.49\textwidth]{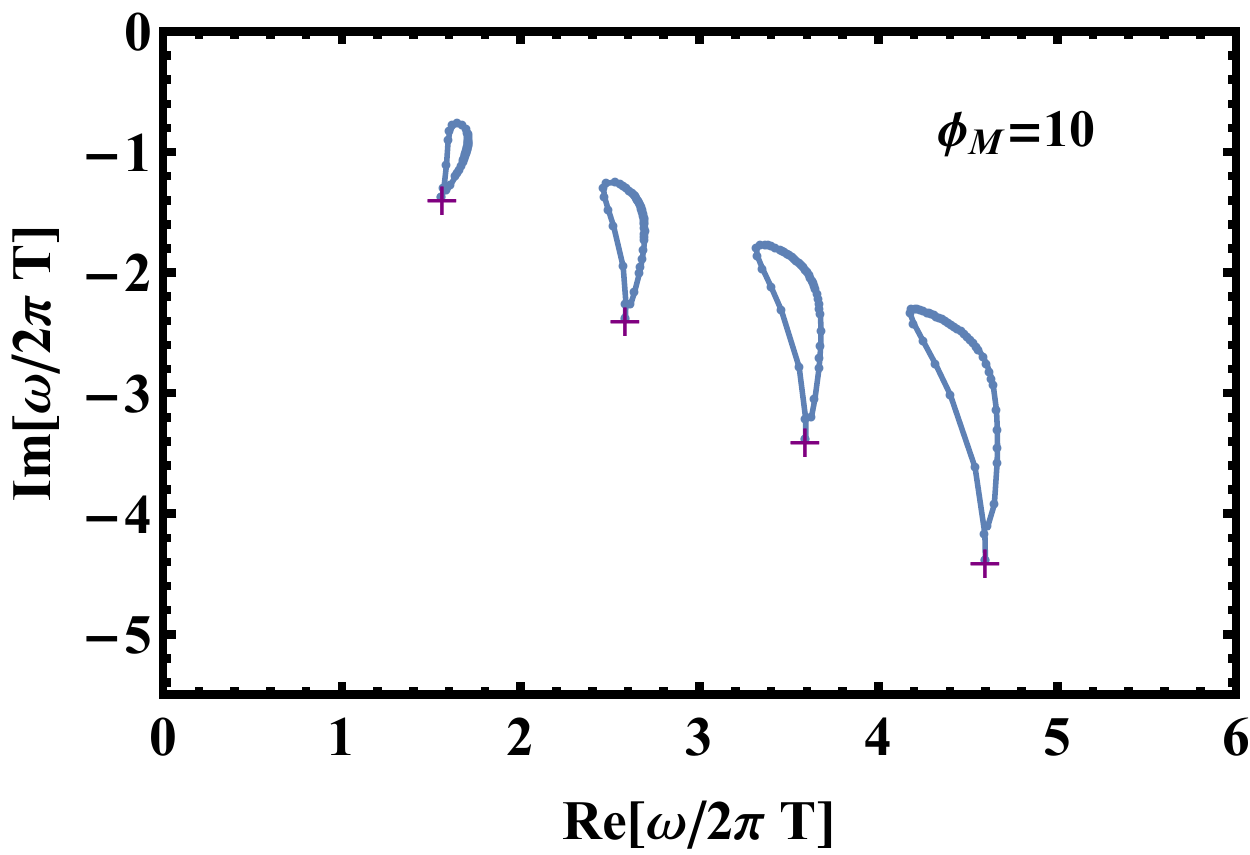} &
\put(140,110){$	\phiM=100$}
\includegraphics[width=.49\textwidth]{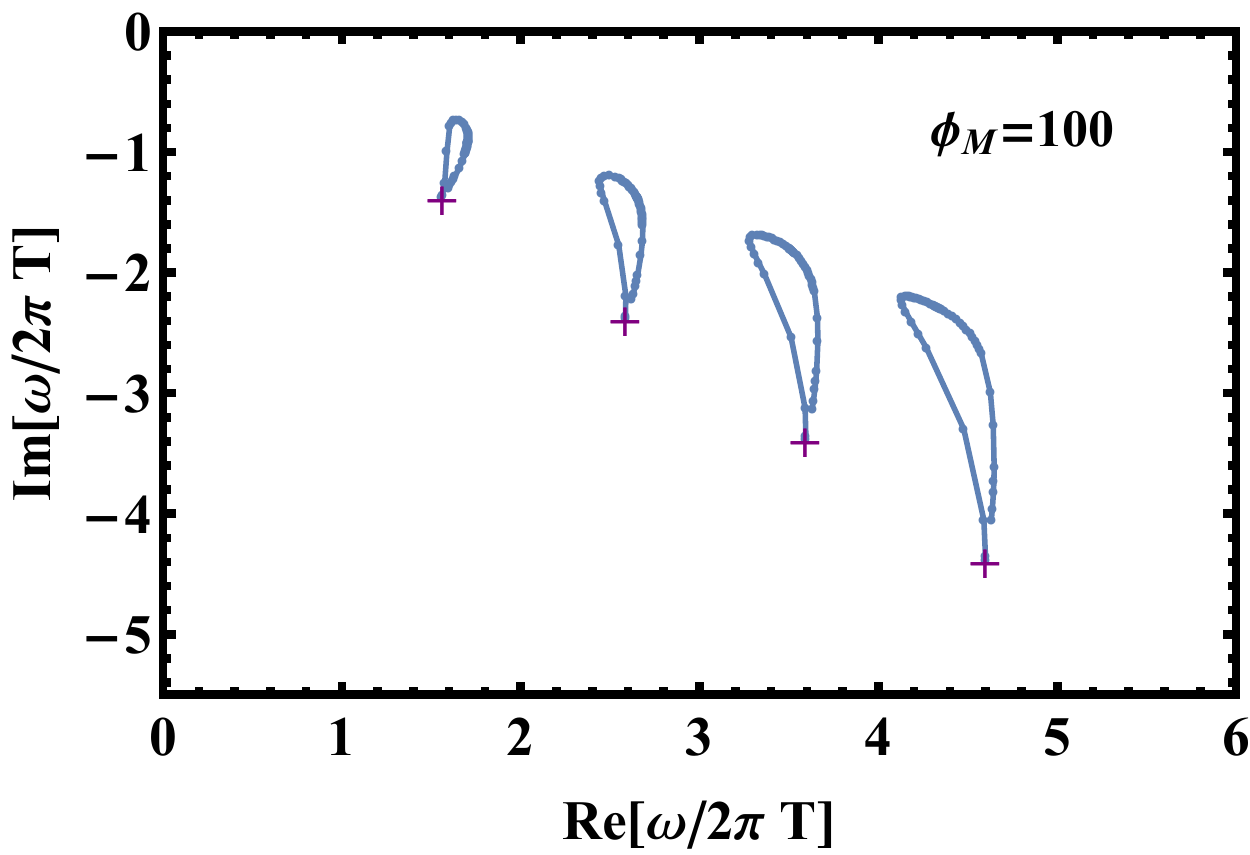}
\end{tabular}
\caption{\label{plot:MetricTower} Complex plane trajectories of the four lowest QNMs of the $\Zm$-channel for different values of $\phiM$. The "+" crosses mark the position of the QNM of thermal $AdS_5$ in this channel. We only show the QNMs with positive real part of the frequency. }
\end{figure}

The QNM frequencies depend on the temperature of  the plasma. As the temperature changes, each of these complex frequencies follows some trajectory in the complex plane.  In \fig{plot:MetricTower} we show these trajectories for the four lowest QNMs of the anisotropic perturbations $\Zm$ for different values of $\phiM$. Each of the points on a given trajectory corresponds to a different value of the temperature. Note that in all panels these trajectories begin and end at the same value, indicated by the crosses, 
"$+$". The reason for this is that the $\Zm$ fluctuations correspond in the gauge theory to fluctuations exclusively of the stress tensor (i.e.~with no contribution of the scalar operator). Since the stress tensor is conserved, its dimension is exactly 4 both at the UV and at the IR fixed points regardless of the value of $\phi_M$. In a CFT, this information of an operator alone would determine the spectrum of the dual QNMs. Since our theory approaches a CFT in the UV and in the IR, the QNMs associated to the pure-stress-tensor fluctuations $\Zm$ approach the same limiting conformal values at high and low temperatures. In contrast,  at intermediate temperatures all the QNM frequencies possess a smaller imaginary part than their conformal counterparts. The magnitude of this deviation depends on the non-conformality of the theory. For $\phiM=1$, when the non-conformal parameter  $I$, \eq{eq:Idef}, is small at all temperatures, the complex plane trajectories of all modes remain close to the conformal value. As $\phiM$ increases the excursion of all modes in the complex plane deviates more from the conformal values. Note, however, that these paths seem to saturate at very high value of the parameter $\phiM$. In particular, even though the change in the number of IR degrees of freedom differs by more than 3 orders of magnitude, the excursion in the complex plane of the simulations with $\phiM=10$ and $\phiM=100$ are very similar. This is in accordance with the small change in non-conformality observed in \fig{plot:I}.

\begin{figure}
\begin{tabular}{cc}
\put(150,110){$	\phiM=1$}
\includegraphics[width=.49\textwidth]{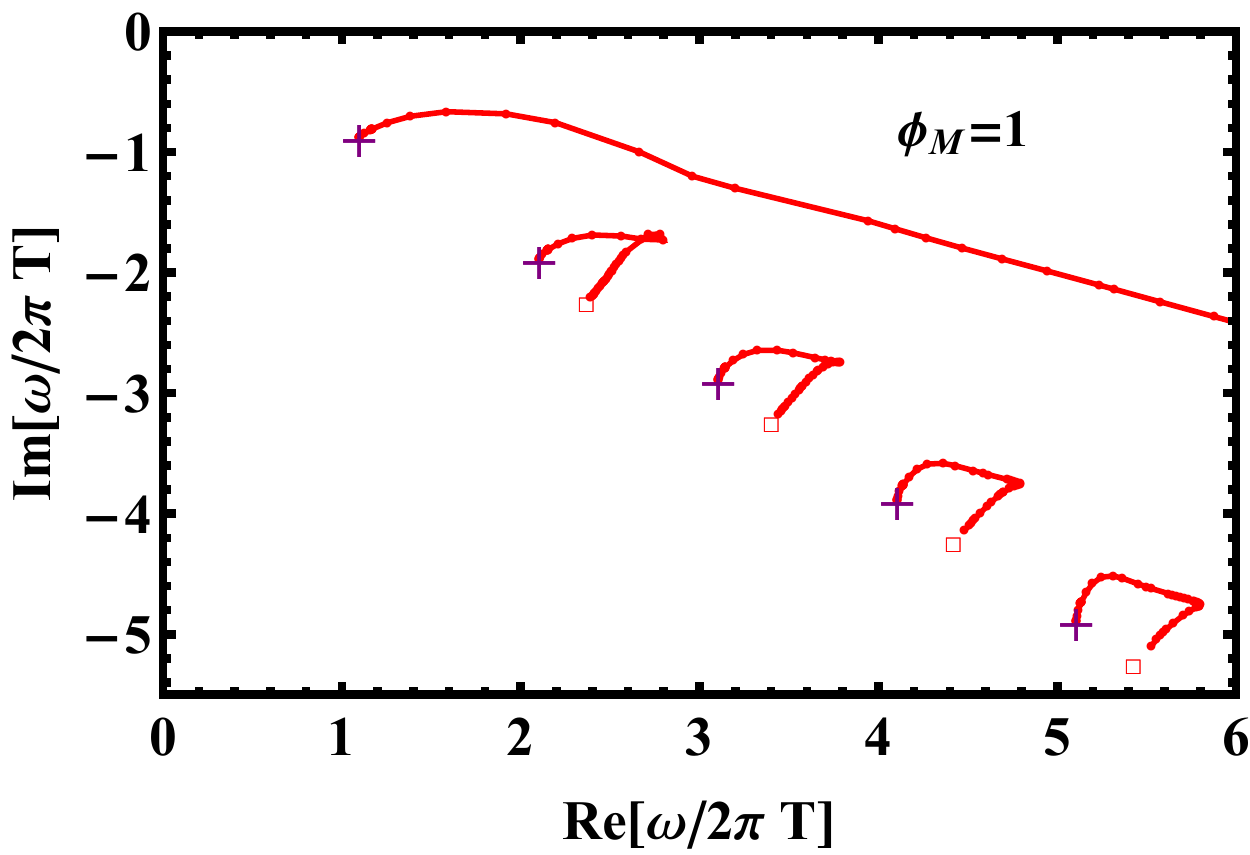} &
\put(150,110){$	\phiM=3$}
\includegraphics[width=.49\textwidth]{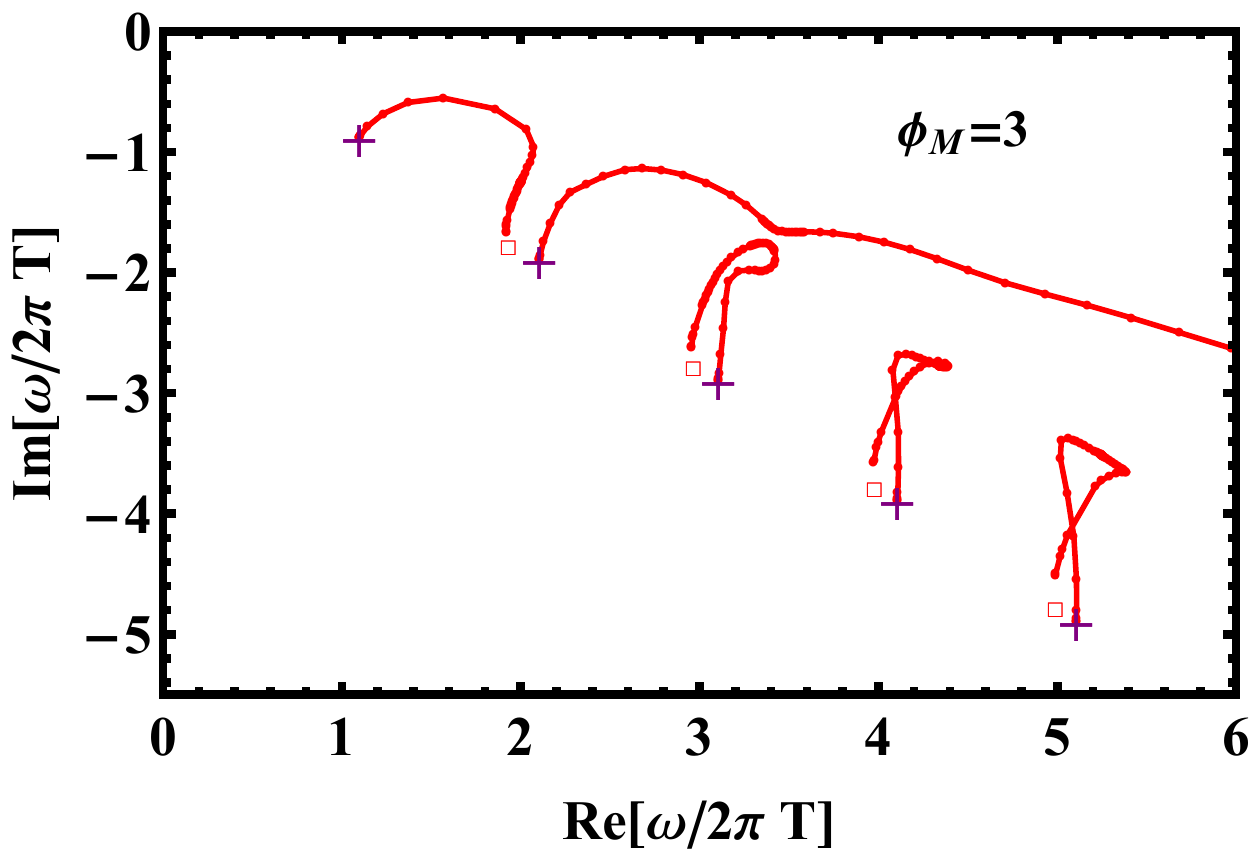}
\\
\put(140,110){$	\phiM=10$}
\includegraphics[width=.49\textwidth]{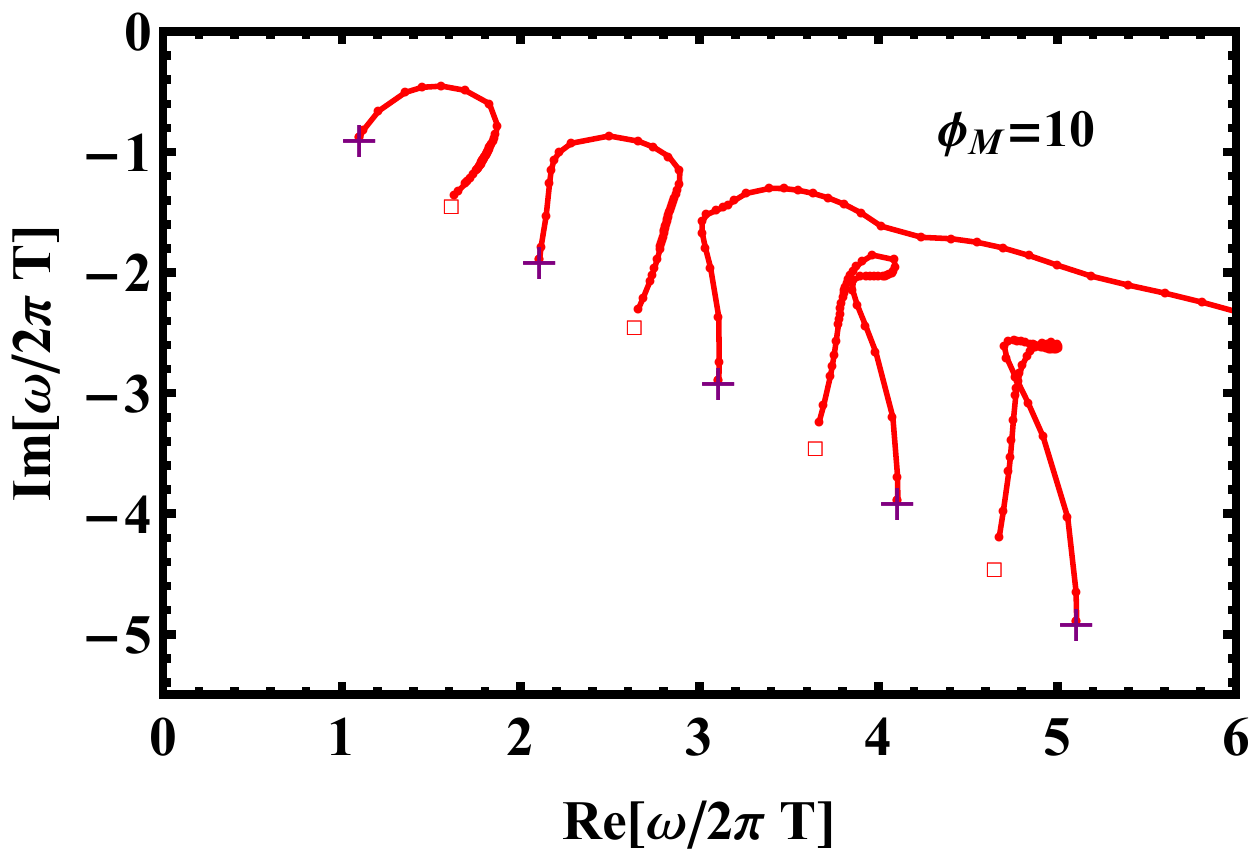} &
\put(140,110){$	\phiM=100$}
\includegraphics[width=.49\textwidth]{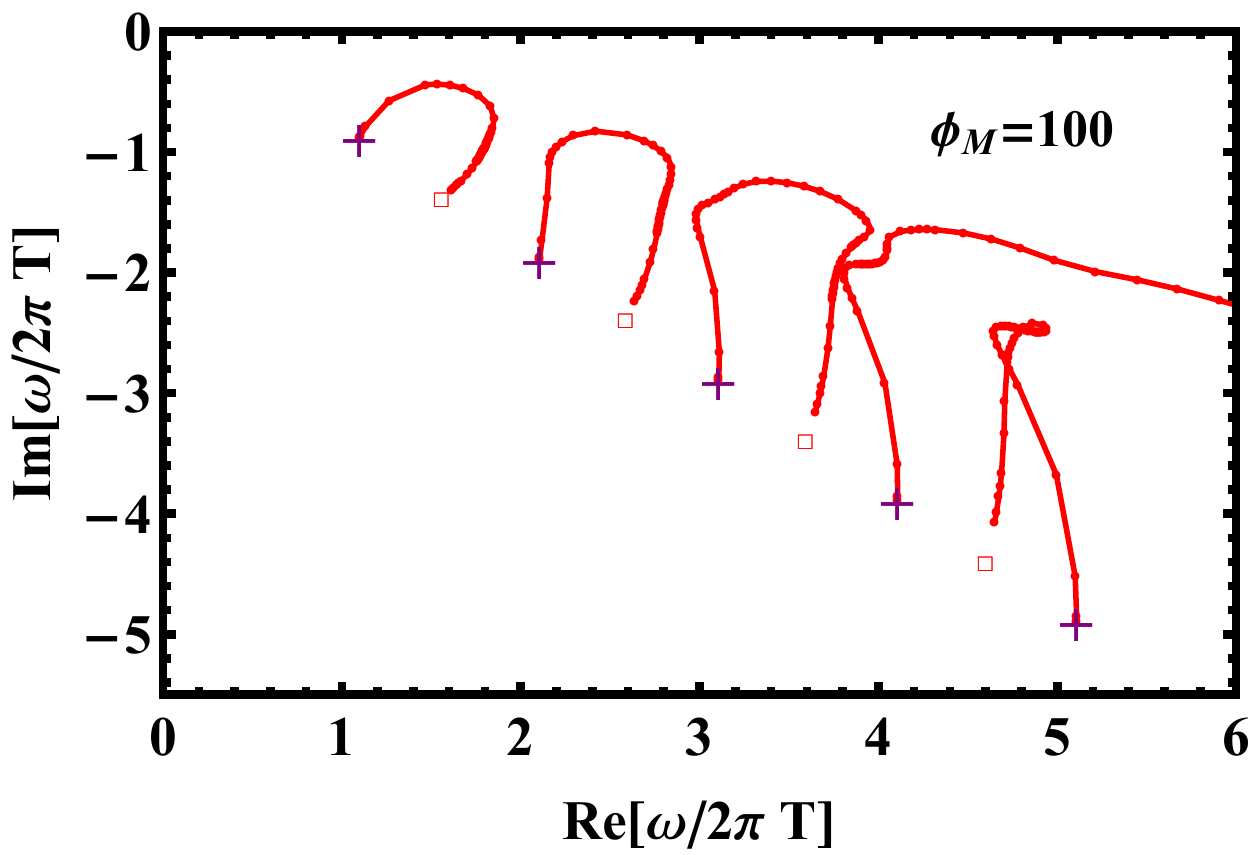}
\end{tabular}
\caption{\label{plot:ScalarTower} Complex-plane trajectories of the four lowest QNMs of the $\Zphi$-channel for different values of $\phiM$. The "+" crosses ("$\Box$" squares) mark the position of the QNM of 
a probe scalar in an AdS black brane background dual to an operator of dimension $\Delta=3$  ($\Delta=\Delta_\mt{IR}$). We only show the QNMs with positive real part of the frequency.}
\end{figure}

In \fig{plot:ScalarTower} we show the complex-plane trajectories of the four lowest QNMs of the bulk mode $\Zphi$ as a function of temperature for different values of $\phiM$. In all panels, the blue "+" crosses show the QNMs of a probe scalar field in an AdS black brane background dual to a CFT scalar operator of dimension 3 \cite{Nunez:2003eq}. Similarly, the red "$\Box$" squares show the QNMs of a probe scalar field in an AdS black brane background dual to an operator of dimension $\Delta_\mt{IR}$ given by \eq{IRdim}. Since this dimension depends on $\phi_M$, the position of the red "$\Box$" squares changes from panel to panel. Based on our discussion of the $\Zm$ QNMs above, one may expect that in the case of $\Zphi$ the trajectories begin at the blue crosses at high temperature and end at the red crosses at low temperature. However, as we can see from \fig{plot:ScalarTower}, the trajectories in this case possess a  more interesting structure.

In the upper left panel of \fig{plot:ScalarTower} we show the trajectories for the $\phiM=1$ potential. For this value, the effective IR mass of the scalar is such that the first two QNMs of the ultraviolet probe scalar are closer to the real axis than the first QNM of the infrared probe scalar. This ordering determines the trajectories of the QNMs as a function of temperature. As shown in the plot, starting from the IR, the lowest  QNM flows towards the closest UV mode in the complex plane, which in this case is the second UV mode. All IR modes follow similar trajectories in such a way that (at least as far as our numerics can resolve)  the $n$-th IR mode flows to the $(n+1)$-th ultraviolet mode.\footnote{We have tested this behaviour for the first 8 QNMs.}
 As a consequence, there are no available IR modes to which the lowest UV mode can flow into. Therefore, this mode decouples at low temperature, flowing deep into the complex imaginary plane. 

For the other values of $\phiM$ displayed in  \fig{plot:ScalarTower}, the positions of the IR and the UV modes alternate in the complex plane, but this does not mean that the flow induced by the temperature is a direct map between these two sets of modes. Even though for the remaining three panels the lowest QNM flows between the lowest modes of the IR and UV theories, in all panels there is always a mode that decouples from the spectrum, although that mode is different for each of the displayed values of $\phi_M$. The origin of this decoupling is that, after a certain mode, the $n$-th IR mode flows to the $(n+1)$-th UV mode, interrupting the trajectory of the $n$-th UV mode. When this happens, we observe a phenomenon similar to level anti-crossing in quantum mechanics. We have checked that for $\phiM=1000$ (not shown) the 
complex-plane trajectories are almost identical to the $\phiM=100$ trajectories displayed in the bottom-right panel of \fig{plot:ScalarTower}. This suggests that the observed structure saturates at large $\phi_M$ and is captured by the $\phiM=100$ plot.

\begin{figure}[h!!!] 
\begin{tabular}{cc}
\put(160,120){$	\phiM=1$}
\includegraphics[width=.49\textwidth]{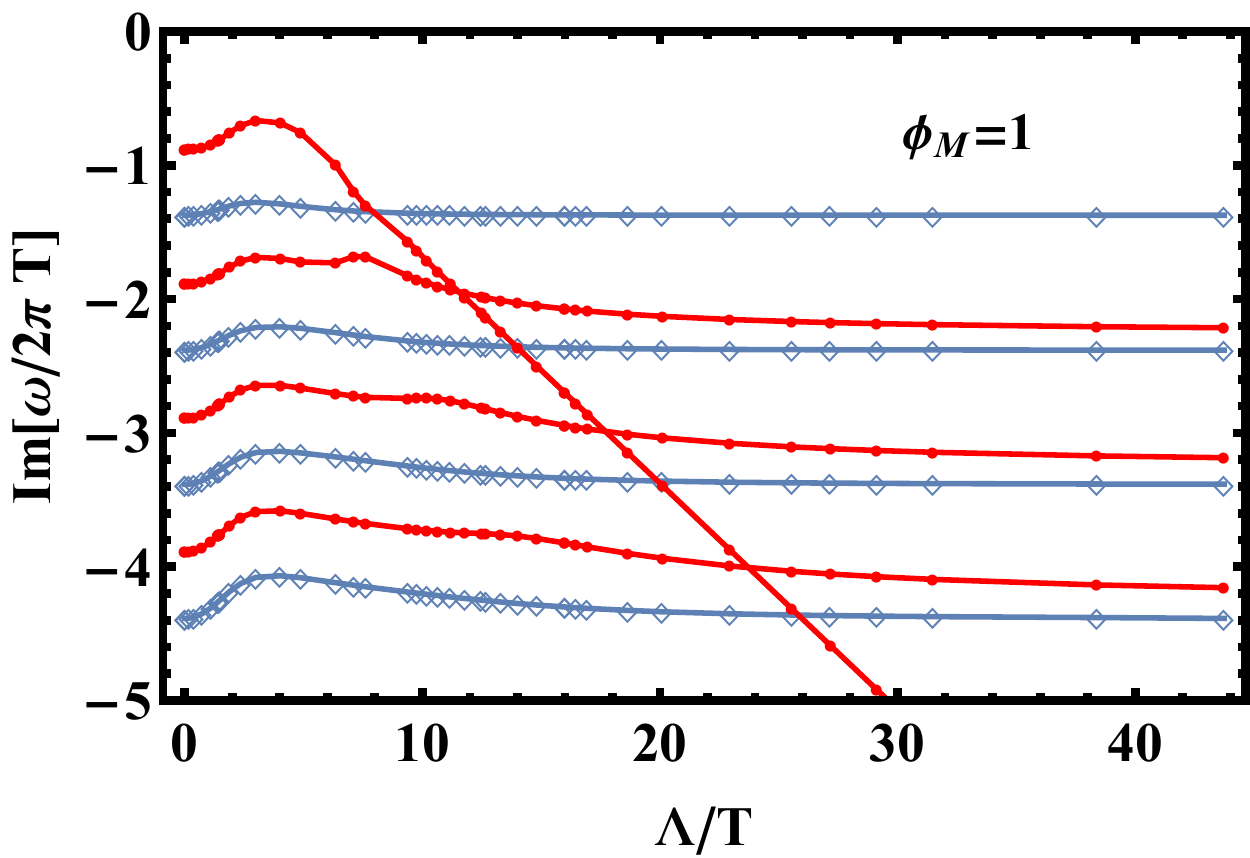}
 &
  \put(160,120){$	\phiM=1$}
 \includegraphics[width=.49\textwidth]{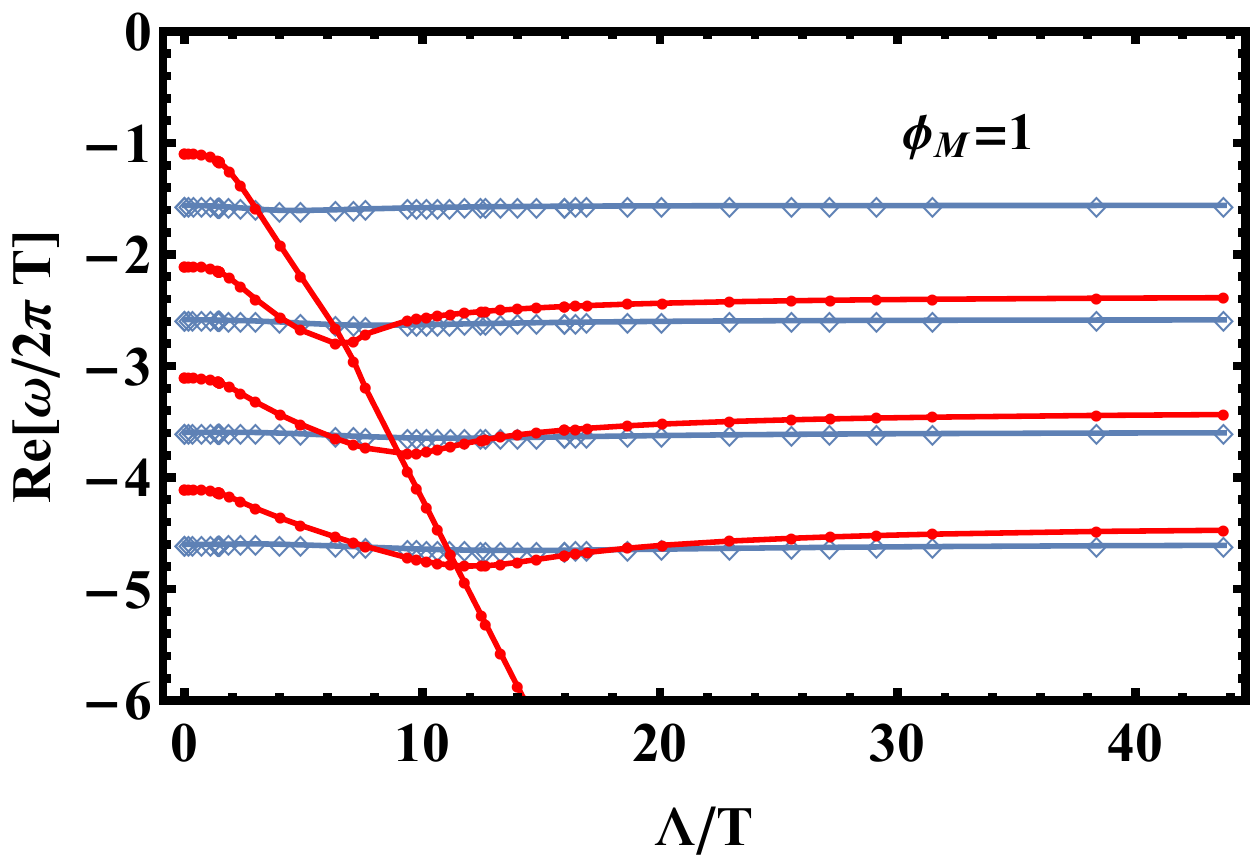}  
 \\
 \put(160,120){$\phiM=3$}
\includegraphics[width=.49\textwidth]{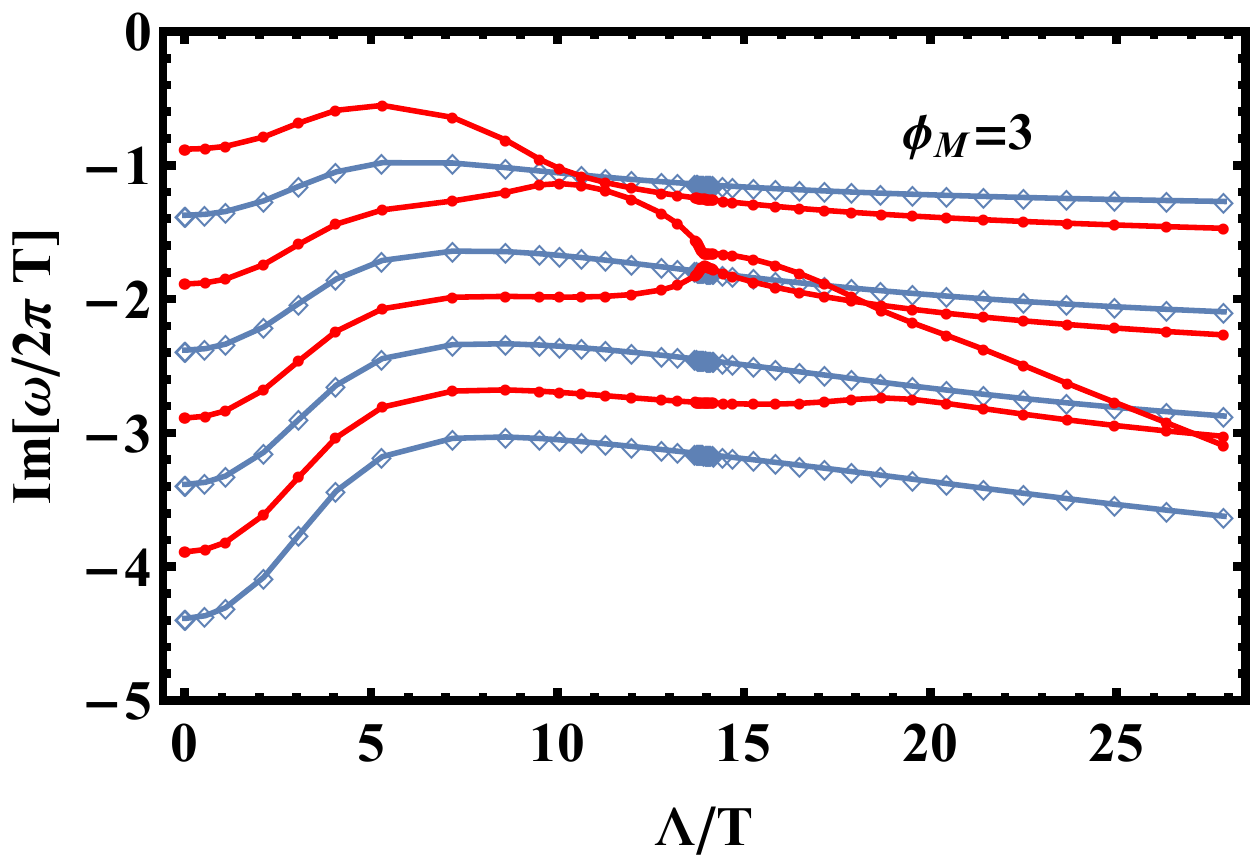}  & 
\put(160,120){$	\phiM=3$}
\includegraphics[width=.49\textwidth]{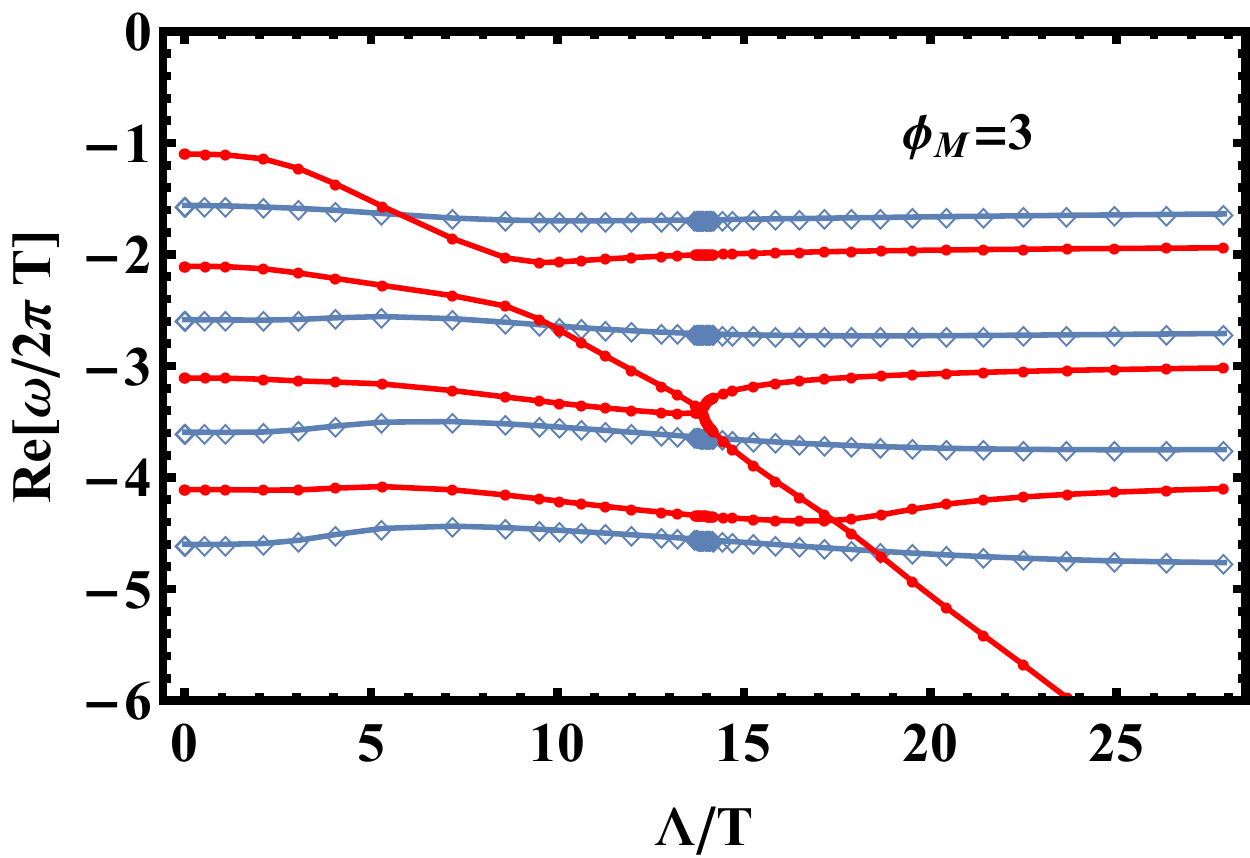} 
\\
\put(160,120){$	\phiM=10$}
\includegraphics[width=.49\textwidth]{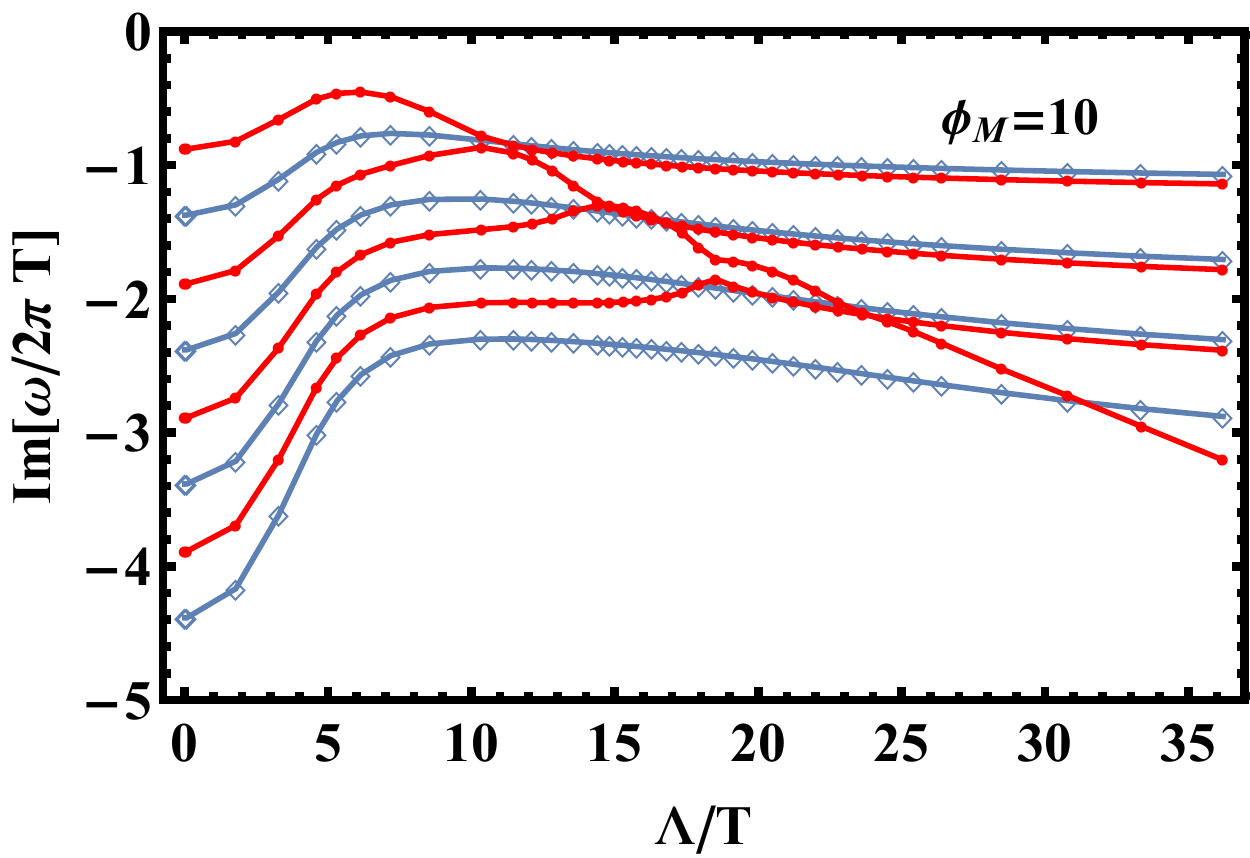} &
\put(160,120){$	\phiM=10$}
\includegraphics[width=.49\textwidth]{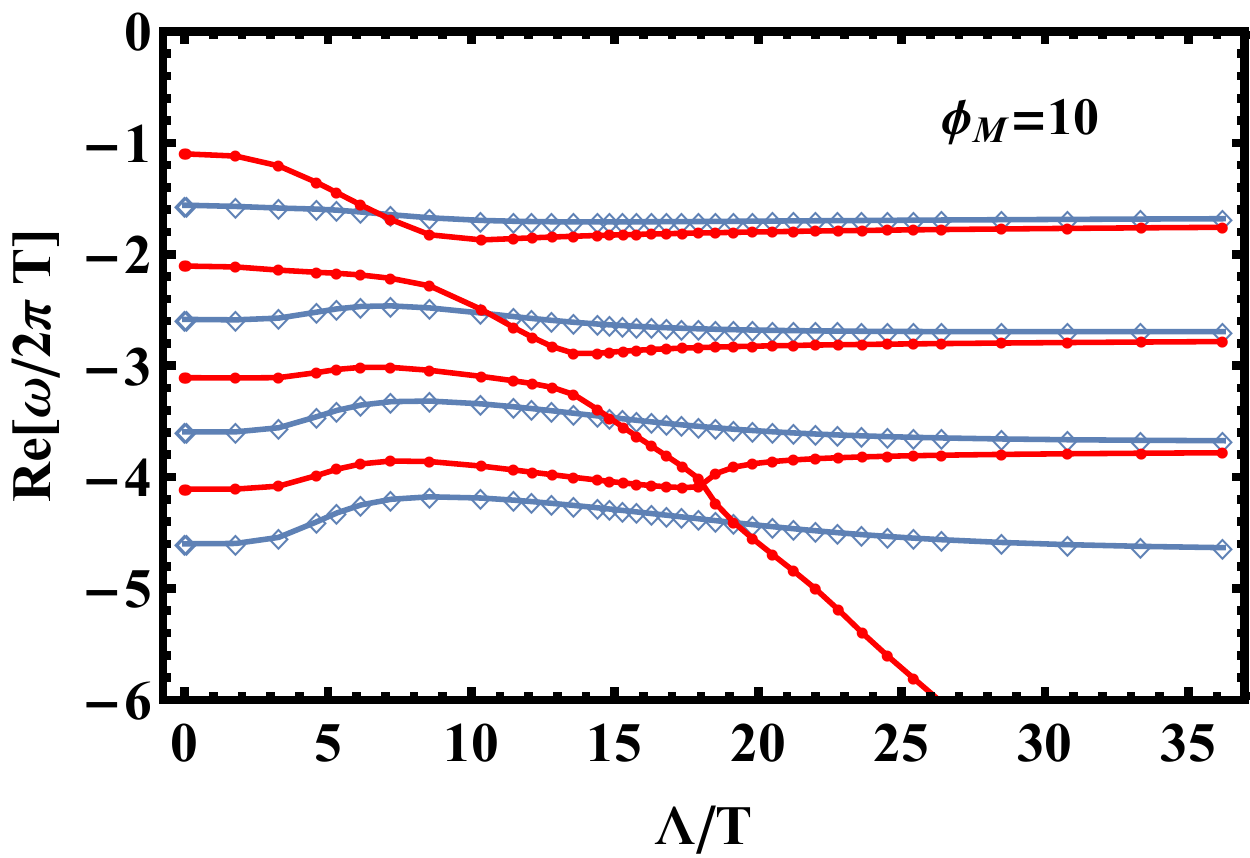}  \\
\put(160,120){$	\phiM=100$}
\includegraphics[width=.49\textwidth]{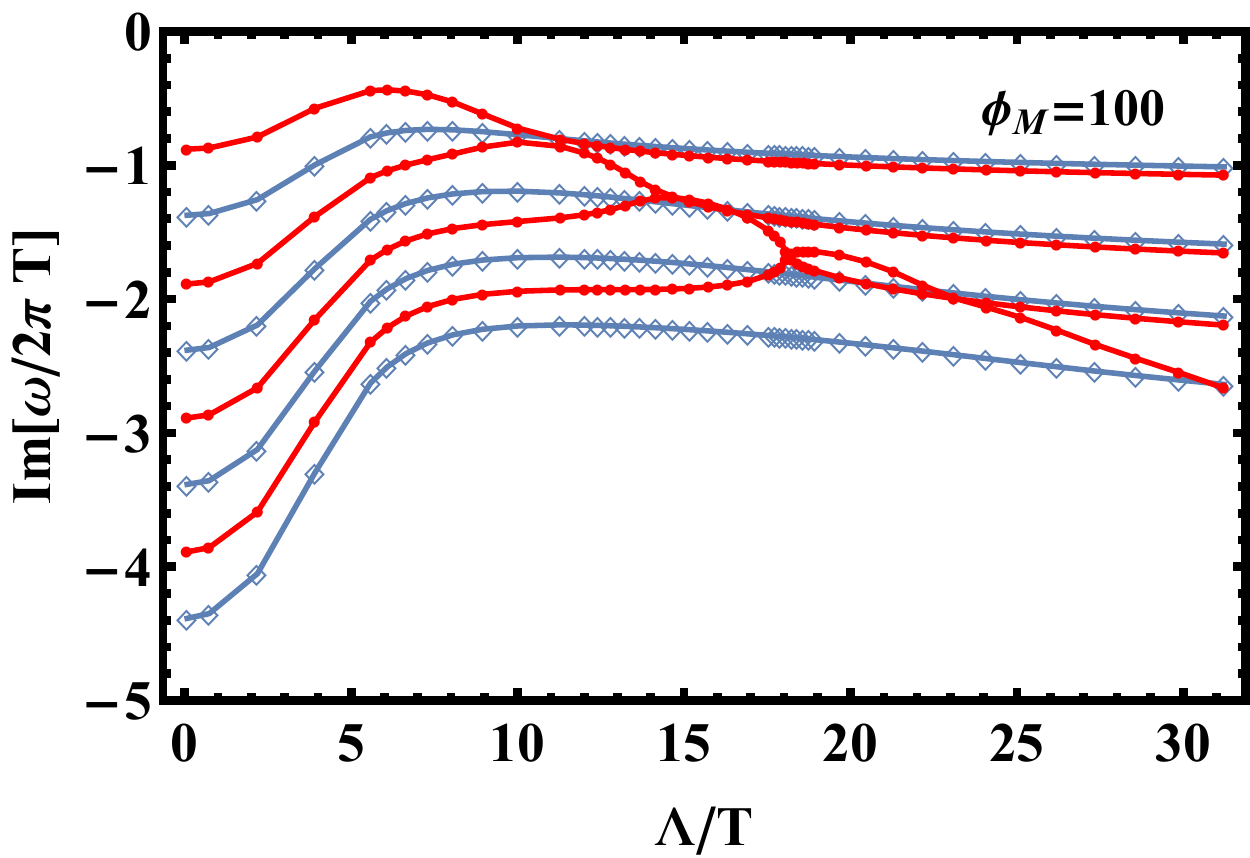} &
\put(160,120){$	\phiM=100$}
\includegraphics[width=.49\textwidth]{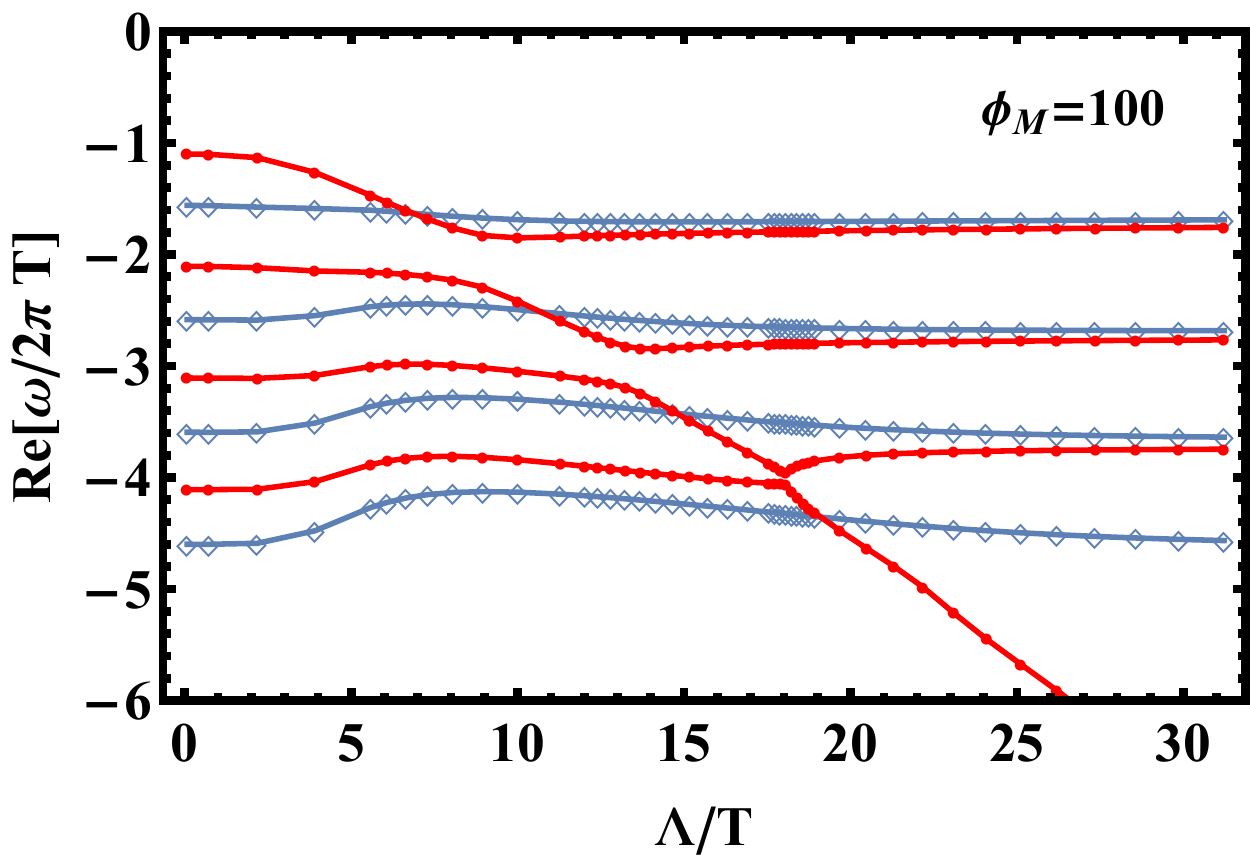}  \\
\end{tabular}
\caption{\label{fig:Tdep} Temperature dependence of the real part (right) and the imaginary part (left) of the four lowest QNMs of the $\Zphi$-channel (red, closed symbols) and $\Zm$-channel (open, blue symbols).}
\end{figure}

In \fig{fig:Tdep} we show the temperature dependence of the imaginary (left) and real (right) parts of the first four quasi-normal frequencies for different values of $\phiM$. 
Each plot shows the QNM of the anisotropic (blue) and bulk (red) channels. 
As already discussed, both of these two sets of modes flow from their values in the UV fixed point to their values in the IR fixed point. 
As shown in the plots, the effective conformal behaviour of the QNMs at high temperature stops when the temperature becomes of order the source $\Lambda$. At higher temperatures, the temperature dependence of both 
the real and imaginary part of the modes is non-trivial, and it reflects the intricate trajectories in the complex plane displayed in \fig{plot:ScalarTower} and   \fig{plot:MetricTower}. These plots also show explicitly how the disappearance of one bulk QNM occurs at low temperature. The fact that this disappearance seems to be linear in all plots in \fig{fig:Tdep} clarifies the temperature dependence of this mode.  The observed constant slope implies that this quasi-normal frequency becomes temperature-independent  at low temperature (we have explicitly checked this) and therefore it decouples from the IR theory. 

As a final remark, we note that the numerical results displayed in \fig{fig:Tdep} allow us to compare the magnitude of the different modes at the same temperature. As we will discuss in more detail in the next section, the imaginary part of the quasi-normal frequencies is related to the relaxation back to equilibrium of small plasma perturbations. It is interesting to note that the ordering of the imaginary parts of the anisotropic and bulk modes changes with temperature: while at high temperatures the imaginary part of the lowest bulk mode is smaller than that of the anisotropic mode, at low  temperatures this order is reversed. This crossing of the imaginary parts of the lowest modes is present for all values of $\phiM$. Nevertheless, 
at $\phiM=1$ this effect is much more prominent, since for this $\phiM$ the disappearing QNM is the lowest bulk mode at high temperature. In the next section we will discuss the consequences of this behaviour.

\section{Discussion}
\label{sec:conclussions}

\begin{figure}
\includegraphics[width=.49\textwidth]{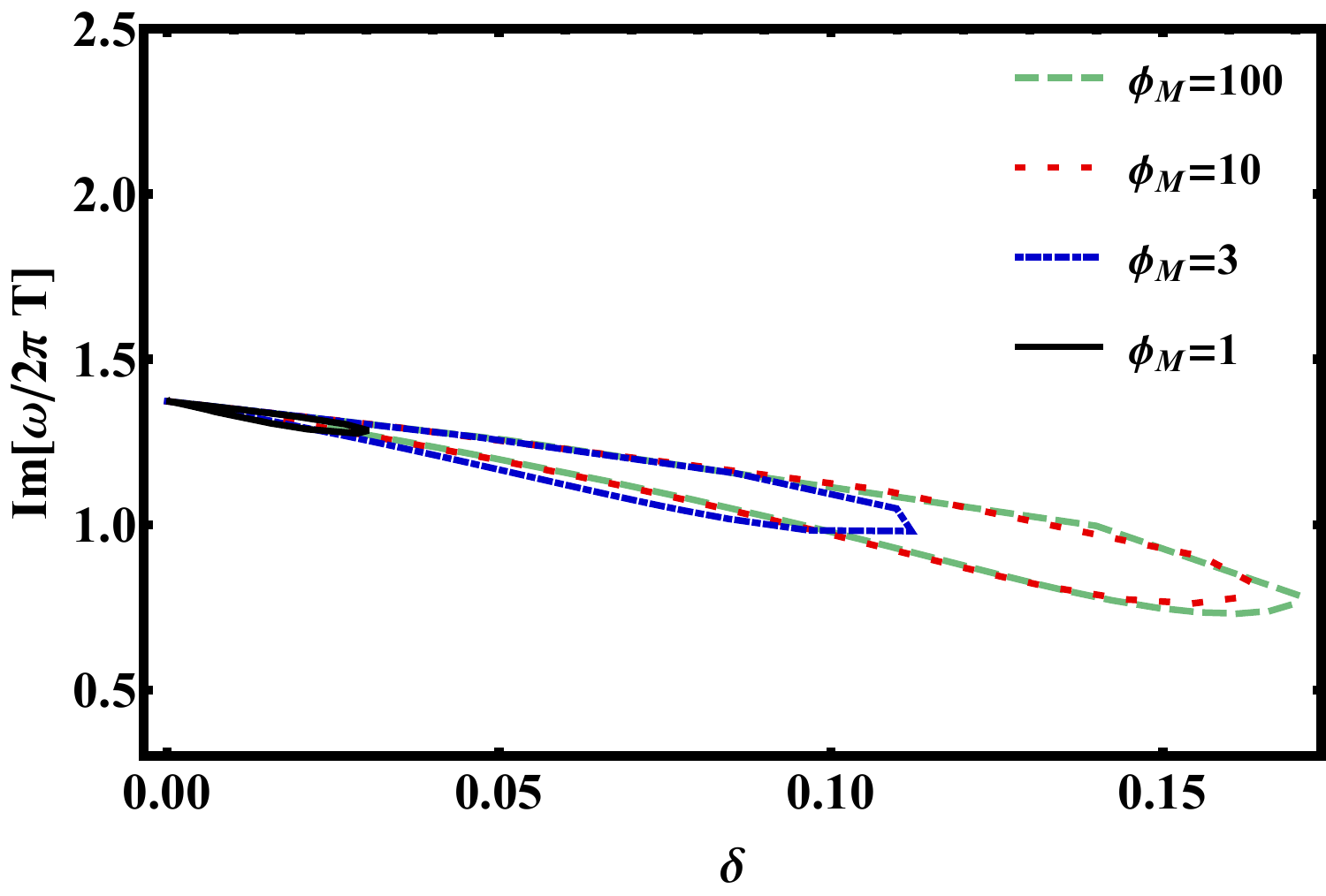}
\includegraphics[width=.49\textwidth]{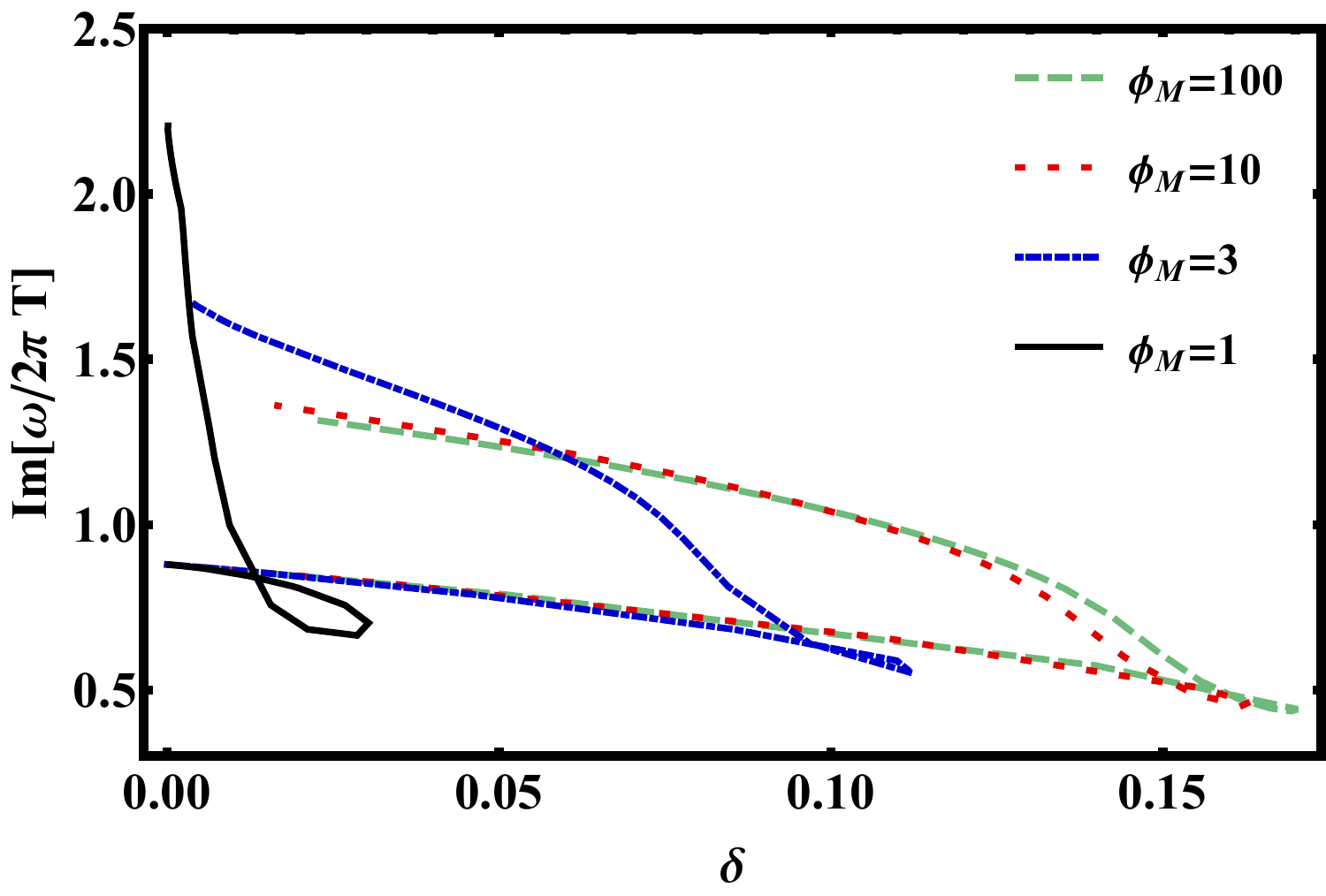}
\caption{ \label{fig:csW} Dependence of the imaginary part of the lowest quasi-normal anisotropic (left) and bulk (right) modes on the speed of sound for different potentials. $\delta=1/3-c_s^2$.}
\end{figure}

The behaviour of the QNM with smallest imaginary part, dubbed the lowest QNM, is particularly relevant for understanding the off-equilibrium dynamics of theories with a gravity dual. 
At non-zero spatial momentum, the lowest QNM of metric perturbations is dual to hydrodynamic excitations of the dual theory.  However, 
in the zero-spatial momentum limit we have considered, the residues of  the hydrodynamic poles  vanish.\footnote{We thank A. Starinets for clarifying this point to us.} In this limit the relaxation back to equilibrium is controlled by the QNM frequencies, with the longest-lived excitation corresponding to the lowest QNM. We will refer to the inverses of the imaginary parts of the frequencies of the lowest QNMs in the different channels as relaxation times.  In the non-conformal theory that we have studied, these important time scales have a very interesting behaviour. 
 
As shown in  \fig{fig:Tdep}, the relaxation time associated to the anisotropic and bulk channels have a non-trivial temperature dependence, as a consequence of non-conformality. To best understand the origin of this temperature dependence, 
 following \cite{Janik:2015waa,Buchel:2015ofa}  in \fig{fig:csW} we show the 
imaginary part of the lowest QNM as a function of $\delta=1/3-c_s^2$ for the anisotropic channel (left) and the bulk channel (right)  for different values of $\phiM$. 
For the anisotropic channel, most of  this dependence may be understood as a consequence of the change of the speed of sound, similarly to the holographic constructions analysed in  \cite{Janik:2015waa,Buchel:2015ofa}. Although the inverse relaxation time 
is not just a common function of $c_s$ for all models, up to small corrections a simple linear dependence of the imaginary part of the lowest anisotropic mode on $\delta=1/3-c_s^2$ provides a good estimate for the relaxation time in this channel.
This simple approximate scaling does not work  in the bulk channel, as shown in the right panel of  \fig{fig:csW}. Unlike the anisotropic channel, the relaxation time is influenced significantly by the change in the scaling dimension of the scalar operator in the high and low
temperature phases, which enters only indirectly into thermodynamic properties such as $c_s$. 
Therefore, the relaxation of strongly coupled gauge theories is, in general, not just controlled by thermodynamic properties, but additional 
microscopic dynamics of the theory may also be important to understand this complicated process. 

The different behaviour of these time scales  reflects the fact that  the way in which the system relaxes depends on the way it is excited. 
To focus the discussion, we will restrict ourselves to generic excitations of the 
stress  tensor of the system. Since in a CFT the trace of the stress tensor vanishes by symmetry, in a CFT this trace cannot be affected by fluctuations of the bulk mode. Since in addition the bulk mode is isotropic, it follows that the stress tensor itself in a CFT cannot be affected by the bulk mode.  Because of this decoupling, the relaxation of  small excitations of the stress tensor
is controlled solely by the lowest mode of the anisotropic channel, given by the $\delta=0$ intercept of  \fig{fig:csW} (left). 
In a non-conformal theory, however, this decoupling does not occur. Because of non-conformality, the fluctuations of the stress tensor and of the operator $\O$ mix. As an example, note that small isotropic variations of the pressure of the system, which at finite momentum are part of the sound channel,  excite the bulk mode, as it can be easily inferred from the Ward identity \eq{eq:TTrace}. 
More generally, the variation of the stress tensor associated to the two fluctuating channels \eqq{considered1}-\eqq{considered2} is given by 
\be
\Delta \epsilon &=&0, \\
\Delta p_z&=&\frac{1}{3} \Lambda^4 \left(\Zphi^{(3)}+4 \Zm^{(4)} \right)\,,
\\
\Delta p_\perp&=&\frac{1}{3} \Lambda^4 \left(\Zphi^{(3)}-2 \Zm^{(4)} \right)\,,
\\
\Delta \left<\O\right>&=&  \Lambda^3 \Zphi^{(3)} \,,
\ee
where $\Delta p_z$ and $\Delta p_\perp$ are the diagonal components of the stress tensor along the direction of the anisotropic perturbation and perpendicular to it, and $\Zphi^{(3)}$ and $\Zm^{(4)}$ are the normalisable modes of the perturbations.
These expressions show explicitly how both channels affect the dynamics of the pressure, while only the bulk channel affects the expectation value of the scalar operator.
As a consequence, the relaxation of the stress tensor of the system  will be dominated by the mode with the smallest imaginary part of the 
two sets of towers displayed in \fig{fig:Tdep}. As it can be seen in this plot, for all values of $\phiM$, relaxation is dominated by different modes at high and low temperatures. 
The competition between these two channels implies that the relaxation dynamics in our family of holographic models follows different paths at high and low temperatures.
 
The contributions of the anisotropic and the bulk modes to the stress tensor codify two different physical processes. As explained above, the anisotropic mode  controls anisotropic perturbations of the pressure that leave unaffected the energy density, the expectation value of the scalar operator, the average pressure and the trace of the stress tensor. The bulk mode controls fluctuations that change the three pressures in an isotropic way and at the same time modify the expectation value of the scalar operator and the relation between the energy density and pressure given by the equation of state.  
%Both these sets of excitations take the system out of equilibrium, and, therefore, at sufficiently late times after the initial disturbance the system relaxes.  This relaxation process is controlled by the QNMs.  
The relaxation of a generic small stress tensor disturbance therefore requires two distinct process: the ``isotropisation'' of the system, which amounts to equating the diagonal spatial components of the stress tensor (pressures); and the 
``EoSization''  of the system, with which  we only refer to the process by which the trace of the stress tensor attains its equilibrium value. We have carefully defined these two terms to avoid any possible confusion with ``thermalization'', namely the process by which a system reaches perfect thermal equilibrium. 

Consider first the case in which the bulk mode dominates the relaxation process, meaning that its associated lowest QNM decays faster than that associated to the anisotropic mode. In this case the system first relaxes the trace of the stress tensor, such that the pressures of the system no longer fluctuate independently,  and only later equates the value of all the pressures to one another. In other words, the system first EoSizes and subsequently isotropizes. This is the behaviour  of the holographic models at small values of $\phiM$, such as $\phiM=1, 3$,   at low temperatures. Since in CFTs the trace of the stress tensor is fixed, this relaxation path is very similar to that in CFTs. 

In contrast, consider now the opposite case in which relaxation is 
dominated by the anisotropic mode, meaning that its associated lowest QNM decays faster than that associated to the bulk mode. In this case the pressure of the system is first isotropized to a value that is not related to the energy density through the equation of state, and only later the subsequent dynamics of this isotropic stress tensor relaxes this value of the pressure to that dictated by the equation of state. At high temperatures, this is the path to equilibration followed by our models,  which differs qualitatively from the conformal case.\footnote{Note that the right-hand side of \eqq{eq:TTrace0} may suggest that $\Lambda \left< \O \right>$ must be large in order to cause a significant violation of the equation of state, thus in possible conflict with the linear approximation. It would be interesting to explore this in a non-linear calculation.} 

Finally, when the two relaxation times are comparable, as it is the case in the low temperature regime for large values of $\phiM=10, 100$,  both of these processes occur simultaneously.

Our calculations are done at zero spatial momentum. At non-zero $k$ the analysis is more complicated because the anisotropic mode splits into the shear, the tensor and the sound modes, and the latter mixes with the bulk mode. Nevertheless, in the coupled bulk-sound system it is still possible to distinguish between those excitations that change the trace of the stress tensor and those that do not. These coupled dynamics will of course modify the EoSization and the isotropization times that we have computed. However, by continuity this modification must be small for small $k$. Since the QNM frequencies are parametrically of order $T$, we therefore expect that their ordering will remain the same provided  $k \ll T$. 

Although the analysis of QNMs can only provide definite answers for the fate of small perturbations off-equilibrium, the rich structure exhibited in this relaxation process has implications for the dynamics of initial configurations that are far off-equilibrium. As we mentioned above, the numerical analyses of collisions in $\mathcal{N}=4$ SYM yield hydrodynamisation times that are comparable to the relaxation times obtained via a QNM analysis. While the microscopic explanation of this observation is not understood, this experience has led the authors of \cite{Buchel:2015saa} to suggest  that the hydrodynamisation of non-conformal theories is basically controlled by the temperature of hydrodynamisation, with small (non-parametric) differences with respect to  the conformal case. Following this reasoning, we may estimate how much longer the hydrodynamisation can be in the  family of theories that we have studied. 
Given the mixing of the bulk and anisotropic modes, this longest relaxation is given by the absolute  minimum of the (negative) imaginary part of the QNM sets which,  as shown in \fig{fig:Tdep}, is always controlled by the bulk mode. Comparing with the relaxation of conformal theories $\tau_{\rm conf}=0.73/2\pi T$, this maximal relaxation is $\tau_{\rm max}/\tau_{\rm conf}= 2.1, \, 2.5,\, 3.0,\, \,3.15$ for $\phiM=1,\, 3,\, 10,\, 100$. These maxima occur at $T/\Lambda=0.33,\, 0.19,\, 0.16,\, 0.16$ for each model respectively.   It would be interesting to test explicitly whether the connection with the linearised analysis persists in full numerical simulations of shock collisions in our non-conformal backgrounds. In particular, this would allow the study of the impact  of the different relaxation channels on the on-set of hydrodynamic behaviour.

\acknowledgments

We thank A.~Buchel, V.~Cardoso, C.~Eling, N.~Evans, P.~Figueras, G. Horowitz, M.~Heller, T.~Ishii, M.~Kaminski, J.~Probst, P.~Romatschke, U.~Sperhake, A.~Starinets, and J.~Tarrio  for discussions. We specially thank A.~Ficnar for explaining to us  the spectral method used to determine the QNMs.  The work of MA has been supported by a Marie Skodowska-Curie Individual Fellowship of the European Commission's Horizon 2020 Programme under contract number 658574 FastTh. JCS is a Royal Society University Research Fellow. JCS was  also supported by a Ram\'on~y~Cajal fellowship,  by  the  Marie Curie Career Integration Grant FP7-PEOPLE-2012-GIG-333786 and by the Spanish MINECO through grant FPA2013-40360-ERC.  CFS and DS acknowledge the support from contracts AYA-2010-15709 (Spanish Ministry of Science and Innovation, MICINN) and ESP2013-47637-P (Spanish Ministry of Economy and Competitivity of Spain, MINECO). IP would like to thank the University of Barcelona for the hospitality during the initial stages of this work. DS acknowledges support via a FPI doctoral contract BES-2012-057909 from MINECO. 
We also acknowledge funding from grants MEC FPA2013-46570-C2-1-P, MEC FPA2013-46570-C2-2-P, MDM-2014-0369 of ICCUB, 2014-SGR-104, 2014-SGR-1474, CPAN CSD2007-00042 Consolider-Ingenio 2010, and ERC Starting Grant HoloLHC-306605.

\clearpage

%Bibtex form
\bibliography{NCTT}

\providecommand{\href}[2]{#2}\begingroup\raggedright\begin{thebibliography}{10}

\bibitem{CasalderreySolana:2011us}
J.~Casalderrey-Solana, H.~Liu, D.~Mateos, K.~Rajagopal and U.~A. Wiedemann,
  \emph{{Gauge/String Duality, Hot QCD and Heavy Ion Collisions}},
  \href{http://arxiv.org/abs/1101.0618}{{\tt 1101.0618}}.

\bibitem{Heller:2011ju}
M.~P. Heller, R.~A. Janik and P.~Witaszczyk, \emph{{The characteristics of
  thermalization of boost-invariant plasma from holography}},
  \href{http://dx.doi.org/10.1103/PhysRevLett.108.201602}{\emph{Phys. Rev.
  Lett.} {\bf 108} (2012) 201602}, [\href{http://arxiv.org/abs/1103.3452}{{\tt
  1103.3452}}].

\bibitem{Chesler:2009cy}
P.~M. Chesler and L.~G. Yaffe, \emph{{Boost invariant flow, black hole
  formation, and far-from-equilibrium dynamics in N = 4 supersymmetric
  Yang-Mills theory}},
  \href{http://dx.doi.org/10.1103/PhysRevD.82.026006}{\emph{Phys. Rev.} {\bf
  D82} (2010) 026006}, [\href{http://arxiv.org/abs/0906.4426}{{\tt
  0906.4426}}].

\bibitem{Chesler:2015wra}
P.~M. Chesler and L.~G. Yaffe, \emph{{Holography and off-center collisions of
  localized shock waves}},
  \href{http://dx.doi.org/10.1007/JHEP10(2015)070}{\emph{JHEP} {\bf 10} (2015)
  070}, [\href{http://arxiv.org/abs/1501.04644}{{\tt 1501.04644}}].

\bibitem{Chesler:2013lia}
P.~M. Chesler and L.~G. Yaffe, \emph{{Numerical solution of gravitational
  dynamics in asymptotically anti-de Sitter spacetimes}},
  \href{http://dx.doi.org/10.1007/JHEP07(2014)086}{\emph{JHEP} {\bf 07} (2014)
  086}, [\href{http://arxiv.org/abs/1309.1439}{{\tt 1309.1439}}].

\bibitem{Casalderrey-Solana:2013sxa}
J.~Casalderrey-Solana, M.~P. Heller, D.~Mateos and W.~van~der Schee,
  \emph{{Longitudinal Coherence in a Holographic Model of Asymmetric
  Collisions}},
  \href{http://dx.doi.org/10.1103/PhysRevLett.112.221602}{\emph{Phys. Rev.
  Lett.} {\bf 112} (2014) 221602}, [\href{http://arxiv.org/abs/1312.2956}{{\tt
  1312.2956}}].

\bibitem{Casalderrey-Solana:2013aba}
J.~Casalderrey-Solana, M.~P. Heller, D.~Mateos and W.~van~der Schee,
  \emph{{From full stopping to transparency in a holographic model of heavy ion
  collisions}},
  \href{http://dx.doi.org/10.1103/PhysRevLett.111.181601}{\emph{Phys. Rev.
  Lett.} {\bf 111} (2013) 181601}, [\href{http://arxiv.org/abs/1305.4919}{{\tt
  1305.4919}}].

\bibitem{Heller:2014wfa}
M.~P. Heller, R.~A. Janik, M.~Spali?ski and P.~Witaszczyk, \emph{{Coupling
  hydrodynamics to nonequilibrium degrees of freedom in strongly interacting
  quark-gluon plasma}},
  \href{http://dx.doi.org/10.1103/PhysRevLett.113.261601}{\emph{Phys. Rev.
  Lett.} {\bf 113} (2014) 261601}, [\href{http://arxiv.org/abs/1409.5087}{{\tt
  1409.5087}}].

\bibitem{Kurkela:2015qoa}
A.~Kurkela and Y.~Zhu, \emph{{Isotropization and hydrodynamization in weakly
  coupled heavy-ion collisions}},
  \href{http://dx.doi.org/10.1103/PhysRevLett.115.182301}{\emph{Phys. Rev.
  Lett.} {\bf 115} (2015) 182301}, [\href{http://arxiv.org/abs/1506.06647}{{\tt
  1506.06647}}].

\bibitem{Ackermann:2000tr}
{\scshape STAR} collaboration, K.~H. Ackermann et~al., \emph{{Elliptic flow in
  Au + Au collisions at (S(NN))**(1/2) = 130 GeV}},
  \href{http://dx.doi.org/10.1103/PhysRevLett.86.402}{\emph{Phys. Rev. Lett.}
  {\bf 86} (2001) 402--407}, [\href{http://arxiv.org/abs/nucl-ex/0009011}{{\tt
  nucl-ex/0009011}}].

\bibitem{Adler:2003kt}
{\scshape PHENIX} collaboration, S.~S. Adler et~al., \emph{{Elliptic flow of
  identified hadrons in Au+Au collisions at s(NN)**(1/2) = 200-GeV}},
  \href{http://dx.doi.org/10.1103/PhysRevLett.91.182301}{\emph{Phys. Rev.
  Lett.} {\bf 91} (2003) 182301},
  [\href{http://arxiv.org/abs/nucl-ex/0305013}{{\tt nucl-ex/0305013}}].

\bibitem{Back:2004mh}
{\scshape PHOBOS} collaboration, B.~B. Back et~al., \emph{{Centrality and
  pseudorapidity dependence of elliptic flow for charged hadrons in Au+Au
  collisions at s(NN)**(1/2) = 200-GeV}},
  \href{http://dx.doi.org/10.1103/PhysRevC.72.051901}{\emph{Phys. Rev.} {\bf
  C72} (2005) 051901}, [\href{http://arxiv.org/abs/nucl-ex/0407012}{{\tt
  nucl-ex/0407012}}].

\bibitem{ATLAS:2012at}
{\scshape ATLAS} collaboration, G.~Aad et~al., \emph{{Measurement of the
  azimuthal anisotropy for charged particle production in $\sqrt{s_{NN}}=2.76$
  TeV lead-lead collisions with the ATLAS detector}},
  \href{http://dx.doi.org/10.1103/PhysRevC.86.014907}{\emph{Phys. Rev.} {\bf
  C86} (2012) 014907}, [\href{http://arxiv.org/abs/1203.3087}{{\tt
  1203.3087}}].

\bibitem{Chatrchyan:2012ta}
{\scshape CMS} collaboration, S.~Chatrchyan et~al., \emph{{Measurement of the
  elliptic anisotropy of charged particles produced in PbPb collisions at
  $\sqrt{s}_{NN}$=2.76 TeV}},
  \href{http://dx.doi.org/10.1103/PhysRevC.87.014902}{\emph{Phys. Rev.} {\bf
  C87} (2013) 014902}, [\href{http://arxiv.org/abs/1204.1409}{{\tt
  1204.1409}}].

\bibitem{Aamodt:2010pa}
{\scshape ALICE} collaboration, K.~Aamodt et~al., \emph{{Elliptic flow of
  charged particles in Pb-Pb collisions at 2.76 TeV}},
  \href{http://dx.doi.org/10.1103/PhysRevLett.105.252302}{\emph{Phys. Rev.
  Lett.} {\bf 105} (2010) 252302}, [\href{http://arxiv.org/abs/1011.3914}{{\tt
  1011.3914}}].

\bibitem{Adam:2016izf}
{\scshape ALICE} collaboration, J.~Adam et~al., \emph{{Anisotropic flow of
  charged particles in Pb-Pb collisions at $\sqrt{s_{\rm NN}}=5.02$ TeV}},
  \href{http://arxiv.org/abs/1602.01119}{{\tt 1602.01119}}.

\bibitem{Aad:2014lta}
{\scshape ATLAS} collaboration, G.~Aad et~al., \emph{{Measurement of long-range
  pseudorapidity correlations and azimuthal harmonics in $\sqrt{s_{NN}}=5.02$
  TeV proton-lead collisions with the ATLAS detector}},
  \href{http://dx.doi.org/10.1103/PhysRevC.90.044906}{\emph{Phys. Rev.} {\bf
  C90} (2014) 044906}, [\href{http://arxiv.org/abs/1409.1792}{{\tt
  1409.1792}}].

\bibitem{Khachatryan:2015waa}
{\scshape CMS} collaboration, V.~Khachatryan et~al., \emph{{Evidence for
  Collective Multiparticle Correlations in p-Pb Collisions}},
  \href{http://dx.doi.org/10.1103/PhysRevLett.115.012301}{\emph{Phys. Rev.
  Lett.} {\bf 115} (2015) 012301}, [\href{http://arxiv.org/abs/1502.05382}{{\tt
  1502.05382}}].

\bibitem{Abelev:2014mda}
{\scshape ALICE} collaboration, B.~B. Abelev et~al., \emph{{Multiparticle
  azimuthal correlations in p -Pb and Pb-Pb collisions at the CERN Large Hadron
  Collider}}, \href{http://dx.doi.org/10.1103/PhysRevC.90.054901}{\emph{Phys.
  Rev.} {\bf C90} (2014) 054901}, [\href{http://arxiv.org/abs/1406.2474}{{\tt
  1406.2474}}].

\bibitem{Aad:2015gqa}
{\scshape ATLAS} collaboration, G.~Aad et~al., \emph{{Observation of long-range
  elliptic anisotropies in $\sqrt{s}=$13 and 2.76 TeV $pp$ collisions with the
  ATLAS detector}},  \href{http://arxiv.org/abs/1509.04776}{{\tt 1509.04776}}.

\bibitem{Chesler:2016ceu}
P.~M. Chesler, \emph{{How big are the smallest drops of quark-gluon plasma?}},
  \href{http://arxiv.org/abs/1601.01583}{{\tt 1601.01583}}.

\bibitem{Chesler:2015bba}
P.~M. Chesler, \emph{{Colliding shock waves and hydrodynamics in small
  systems}},
  \href{http://dx.doi.org/10.1103/PhysRevLett.115.241602}{\emph{Phys. Rev.
  Lett.} {\bf 115} (2015) 241602}, [\href{http://arxiv.org/abs/1506.02209}{{\tt
  1506.02209}}].

\bibitem{Borsanyi:2013bia}
S.~Borsanyi, Z.~Fodor, C.~Hoelbling, S.~D. Katz, S.~Krieg and K.~K. Szabo,
  \emph{{Full result for the QCD equation of state with 2+1 flavors}},
  \href{http://dx.doi.org/10.1016/j.physletb.2014.01.007}{\emph{Phys. Lett.}
  {\bf B730} (2014) 99--104}, [\href{http://arxiv.org/abs/1309.5258}{{\tt
  1309.5258}}].

\bibitem{Bazavov:2014pvz}
{\scshape HotQCD} collaboration, A.~Bazavov et~al., \emph{{Equation of state in
  ( 2+1 )-flavor QCD}},
  \href{http://dx.doi.org/10.1103/PhysRevD.90.094503}{\emph{Phys. Rev.} {\bf
  D90} (2014) 094503}, [\href{http://arxiv.org/abs/1407.6387}{{\tt
  1407.6387}}].

\bibitem{Ryu:2015vwa}
S.~Ryu, J.~F. Paquet, C.~Shen, G.~S. Denicol, B.~Schenke, S.~Jeon et~al.,
  \emph{{Importance of the Bulk Viscosity of QCD in Ultrarelativistic Heavy-Ion
  Collisions}},
  \href{http://dx.doi.org/10.1103/PhysRevLett.115.132301}{\emph{Phys. Rev.
  Lett.} {\bf 115} (2015) 132301}, [\href{http://arxiv.org/abs/1502.01675}{{\tt
  1502.01675}}].

\bibitem{Bozek:2011if}
P.~Bozek, \emph{{Collective flow in p-Pb and d-Pd collisions at TeV energies}},
  \href{http://dx.doi.org/10.1103/PhysRevC.85.014911}{\emph{Phys. Rev.} {\bf
  C85} (2012) 014911}, [\href{http://arxiv.org/abs/1112.0915}{{\tt
  1112.0915}}].

\bibitem{Schenke:2014zha}
B.~Schenke and R.~Venugopalan, \emph{{Eccentric protons? Sensitivity of flow to
  system size and shape in p+p, p+Pb and Pb+Pb collisions}},
  \href{http://dx.doi.org/10.1103/PhysRevLett.113.102301}{\emph{Phys. Rev.
  Lett.} {\bf 113} (2014) 102301}, [\href{http://arxiv.org/abs/1405.3605}{{\tt
  1405.3605}}].

\bibitem{Habich:2015rtj}
M.~Habich, G.~A. Miller, P.~Romatschke and W.~Xiang, \emph{{A Hydrodynamic
  Study of p+p Collisions at $\sqrt{s}=7$ TeV}},
  \href{http://arxiv.org/abs/1512.05354}{{\tt 1512.05354}}.

\bibitem{Jeon:2015dfa}
S.~Jeon and U.~Heinz, \emph{{Introduction to Hydrodynamics}},
  \href{http://dx.doi.org/10.1142/S0218301315300106}{\emph{Int. J. Mod. Phys.}
  {\bf E24} (2015) 1530010}, [\href{http://arxiv.org/abs/1503.03931}{{\tt
  1503.03931}}].

\bibitem{Heller:2013oxa}
M.~P. Heller, D.~Mateos, W.~van~der Schee and M.~Triana, \emph{{Holographic
  isotropization linearized}},
  \href{http://dx.doi.org/10.1007/JHEP09(2013)026}{\emph{JHEP} {\bf 09} (2013)
  026}, [\href{http://arxiv.org/abs/1304.5172}{{\tt 1304.5172}}].

\bibitem{Janik:2015iry}
R.~A. Janik, J.~Jankowski and H.~Soltanpanahi, \emph{{Non-equilibrium dynamics
  and phase transitions}},  \href{http://arxiv.org/abs/1512.06871}{{\tt
  1512.06871}}.

\bibitem{Buchel:2015saa}
A.~Buchel, M.~P. Heller and R.~C. Myers, \emph{{Equilibration rates in a
  strongly coupled nonconformal quark-gluon plasma}},
  \href{http://dx.doi.org/10.1103/PhysRevLett.114.251601}{\emph{Phys. Rev.
  Lett.} {\bf 114} (2015) 251601}, [\href{http://arxiv.org/abs/1503.07114}{{\tt
  1503.07114}}].

\bibitem{Janik:2015waa}
R.~A. Janik, G.~Plewa, H.~Soltanpanahi and M.~Spalinski, \emph{{Linearized
  nonequilibrium dynamics in nonconformal plasma}},
  \href{http://dx.doi.org/10.1103/PhysRevD.91.126013}{\emph{Phys. Rev.} {\bf
  D91} (2015) 126013}, [\href{http://arxiv.org/abs/1503.07149}{{\tt
  1503.07149}}].

\bibitem{Buchel:2015ofa}
A.~Buchel and A.~Day, \emph{{Universal relaxation in quark-gluon plasma at
  strong coupling}},
  \href{http://dx.doi.org/10.1103/PhysRevD.92.026009}{\emph{Phys. Rev.} {\bf
  D92} (2015) 026009}, [\href{http://arxiv.org/abs/1505.05012}{{\tt
  1505.05012}}].

\bibitem{Rougemont:2015wca}
R.~Rougemont, A.~Ficnar, S.~Finazzo and J.~Noronha, \emph{{Energy loss,
  equilibration, and thermodynamics of a baryon rich strongly coupled
  quark-gluon plasma}},  \href{http://arxiv.org/abs/1507.06556}{{\tt
  1507.06556}}.

\bibitem{Gursoy:2015nza}
U.~Gursoy, M.~Jarvinen and G.~Policastro, \emph{{Late time behavior of
  non-conformal plasmas}},
  \href{http://dx.doi.org/10.1007/JHEP01(2016)134}{\emph{JHEP} {\bf 01} (2016)
  134}, [\href{http://arxiv.org/abs/1507.08628}{{\tt 1507.08628}}].

\bibitem{Gursoy:2016tgf}
U.~Gursoy, A.~Jansen, W.~Sybesma and S.~Vandoren, \emph{{Holographic
  Equilibration of Nonrelativistic Plasmas}},
  \href{http://arxiv.org/abs/1602.01375}{{\tt 1602.01375}}.

\bibitem{Ali-Akbari:2016sms}
M.~Ali-Akbari, F.~Charmchi, H.~Ebrahim and L.~Shahkarami, \emph{{Various
  Time-Scales of Relaxation}},  \href{http://arxiv.org/abs/1602.07903}{{\tt
  1602.07903}}.

\bibitem{Bianchi:2001kw}
M.~Bianchi, D.~Z. Freedman and K.~Skenderis, \emph{{Holographic
  renormalization}},
  \href{http://dx.doi.org/10.1016/S0550-3213(02)00179-7}{\emph{Nucl. Phys.}
  {\bf B631} (2002) 159--194}, [\href{http://arxiv.org/abs/hep-th/0112119}{{\tt
  hep-th/0112119}}].

\bibitem{Girardello:1998pd}
L.~Girardello, M.~Petrini, M.~Porrati and A.~Zaffaroni, \emph{{Novel local CFT
  and exact results on perturbations of N=4 superYang Mills from AdS
  dynamics}},
  \href{http://dx.doi.org/10.1088/1126-6708/1998/12/022}{\emph{JHEP} {\bf 12}
  (1998) 022}, [\href{http://arxiv.org/abs/hep-th/9810126}{{\tt
  hep-th/9810126}}].

\bibitem{Papadimitriou:2004rz}
I.~Papadimitriou and K.~Skenderis, \emph{{Correlation functions in holographic
  RG flows}},
  \href{http://dx.doi.org/10.1088/1126-6708/2004/10/075}{\emph{JHEP} {\bf 10}
  (2004) 075}, [\href{http://arxiv.org/abs/hep-th/0407071}{{\tt
  hep-th/0407071}}].

\bibitem{Megias:2010ku}
E.~Megias, H.~J. Pirner and K.~Veschgini, \emph{{QCD thermodynamics using
  five-dimensional gravity}},
  \href{http://dx.doi.org/10.1103/PhysRevD.83.056003}{\emph{Phys. Rev.} {\bf
  D83} (2011) 056003}, [\href{http://arxiv.org/abs/1009.2953}{{\tt
  1009.2953}}].

\bibitem{Veschgini:2010ws}
K.~Veschgini, E.~Megias and H.~J. Pirner, \emph{{Trouble Finding the Optimal
  AdS/QCD}},
  \href{http://dx.doi.org/10.1016/j.physletb.2011.01.011}{\emph{Phys. Lett.}
  {\bf B696} (2011) 495--498}, [\href{http://arxiv.org/abs/1009.4639}{{\tt
  1009.4639}}].

\bibitem{Gursoy:2008za}
U.~Gursoy, E.~Kiritsis, L.~Mazzanti and F.~Nitti, \emph{{Holography and
  Thermodynamics of 5D Dilaton-gravity}},
  \href{http://dx.doi.org/10.1088/1126-6708/2009/05/033}{\emph{JHEP} {\bf 05}
  (2009) 033}, [\href{http://arxiv.org/abs/0812.0792}{{\tt 0812.0792}}].

\bibitem{Gubser:2008ny}
S.~S. Gubser and A.~Nellore, \emph{{Mimicking the QCD equation of state with a
  dual black hole}},
  \href{http://dx.doi.org/10.1103/PhysRevD.78.086007}{\emph{Phys. Rev.} {\bf
  D78} (2008) 086007}, [\href{http://arxiv.org/abs/0804.0434}{{\tt
  0804.0434}}].

\bibitem{Kovtun:2004de}
P.~Kovtun, D.~T. Son and A.~O. Starinets, \emph{{Viscosity in strongly
  interacting quantum field theories from black hole physics}},
  \href{http://dx.doi.org/10.1103/PhysRevLett.94.111601}{\emph{Phys. Rev.
  Lett.} {\bf 94} (2005) 111601},
  [\href{http://arxiv.org/abs/hep-th/0405231}{{\tt hep-th/0405231}}].

\bibitem{Eling:2011ms}
C.~Eling and Y.~Oz, \emph{{A Novel Formula for Bulk Viscosity from the Null
  Horizon Focusing Equation}},
  \href{http://dx.doi.org/10.1007/JHEP06(2011)007}{\emph{JHEP} {\bf 06} (2011)
  007}, [\href{http://arxiv.org/abs/1103.1657}{{\tt 1103.1657}}].

\bibitem{Gubser:2008sz}
S.~S. Gubser, S.~S. Pufu and F.~D. Rocha, \emph{{Bulk viscosity of strongly
  coupled plasmas with holographic duals}},
  \href{http://dx.doi.org/10.1088/1126-6708/2008/08/085}{\emph{JHEP} {\bf 08}
  (2008) 085}, [\href{http://arxiv.org/abs/0806.0407}{{\tt 0806.0407}}].

\bibitem{Mas:2007ng}
J.~Mas and J.~Tarrio, \emph{{Hydrodynamics from the Dp-brane}},
  \href{http://dx.doi.org/10.1088/1126-6708/2007/05/036}{\emph{JHEP} {\bf 05}
  (2007) 036}, [\href{http://arxiv.org/abs/hep-th/0703093}{{\tt
  hep-th/0703093}}].

\bibitem{Buchel:2011uj}
A.~Buchel, \emph{{Violation of the holographic bulk viscosity bound}},
  \href{http://dx.doi.org/10.1103/PhysRevD.85.066004}{\emph{Phys. Rev.} {\bf
  D85} (2012) 066004}, [\href{http://arxiv.org/abs/1110.0063}{{\tt
  1110.0063}}].

\bibitem{Kovtun:2005ev}
P.~K. Kovtun and A.~O. Starinets, \emph{{Quasinormal modes and holography}},
  \href{http://dx.doi.org/10.1103/PhysRevD.72.086009}{\emph{Phys. Rev.} {\bf
  D72} (2005) 086009}, [\href{http://arxiv.org/abs/hep-th/0506184}{{\tt
  hep-th/0506184}}].

\bibitem{Benincasa:2005iv}
P.~Benincasa, A.~Buchel and A.~O. Starinets, \emph{{Sound waves in strongly
  coupled non-conformal gauge theory plasma}},
  \href{http://dx.doi.org/10.1016/j.nuclphysb.2005.11.005}{\emph{Nucl. Phys.}
  {\bf B733} (2006) 160--187}, [\href{http://arxiv.org/abs/hep-th/0507026}{{\tt
  hep-th/0507026}}].

\bibitem{Nunez:2003eq}
A.~Nunez and A.~O. Starinets, \emph{{AdS / CFT correspondence, quasinormal
  modes, and thermal correlators in N=4 SYM}},
  \href{http://dx.doi.org/10.1103/PhysRevD.67.124013}{\emph{Phys. Rev.} {\bf
  D67} (2003) 124013}, [\href{http://arxiv.org/abs/hep-th/0302026}{{\tt
  hep-th/0302026}}].

\end{thebibliography}\endgroup
\bibliographystyle{JHEP}

%\begin{thebibliography}{99}
% 
%\end{thebibliography}

\end{document}